\documentclass[11pt,a4paper]{article}
\pdfoutput=1
\usepackage{jheppub}

\usepackage{epsfig}
\usepackage{latexsym}
\usepackage{amsfonts}
\usepackage{amsmath}
\usepackage{amsthm}
\usepackage{amssymb}
\usepackage{amsbsy}
\usepackage{multirow}
\usepackage{hyperref}

\newcommand{\RR}{\mathbb{R}} 

\newcommand{\M}{\mathcal{M}}

\def\tr         {{\rm  tr}}
\def\cala         {{\cal A}}

\def\calb         {{\cal B}}

\def\calo         {{\cal O}}
\def\calp         {{\cal P}}

\def\calt         {{\cal T}}

\def\calw         {{\cal W}}
\newsavebox{\uuunit}
\sbox{\uuunit}
    {\setlength{\unitlength}{0.825em}
     \begin{picture}(0.6,0.7)
        \thinlines
        \put(0,0){\line(1,0){0.5}}
        \put(0.15,0){\line(0,1){0.7}}
        \put(0.35,0){\line(0,1){0.8}}
       \multiput(0.3,0.8)(-0.04,-0.02){12}{\rule{0.5pt}{0.5pt}}
     \end {picture}}

\def\be{\begin{equation}}
\def\ee{\end{equation}}
\def\bea{\begin{eqnarray}}
\def\eea{\end{eqnarray}}

\def\a{\alpha}

\def\h{\eta}

\def\d{\delta}
\def\e{\epsilon}

\def\l{\lambda}
\def\L{\Lambda}

\def\f{\phi}

\def\m{\mu}
\def\n{\nu}
\def\o{\omega}

\def\p{\pi}

\def\s{\sigma}

\def\sF{{{ F}\!\!\!\!\hskip.8pt\hbox{\raise1pt\hbox{/}}\,}}
\def\som{{{ \omega}\!\!\!\!\hskip.8pt\hbox{\raise1pt\hbox{/}}\,}}
\def\sJ{{{\rm J}\!\!\!\!\hskip.8pt\hbox{\raise1pt\hbox{/}}\,}}

\def\pa{\partial}

\def\to{\rightarrow}
\def\nonu{\nonumber \\{}}
\def\half{{1 \over 2}}

\theoremstyle{definition}
\newtheorem*{remark}{Remark}
\newcommand{\rem}{\begin{remark}}
\newcommand{\erem}{\end{remark}}

\newcommand{\qq}[1]{``#1''}

\newcommand{\slin}{ \begin{center}
\noindent\rule[0.5ex]{9cm}{1.5pt}
\end{center} }

\newcommand{\im}{\mathrm{i}}
\newcommand{\dif}{\mathrm{d}}
\DeclareMathOperator{\ex}{e}

\title{Multi-centered higher spin solutions from $W_N$ conformal blocks}
\author[a,b]{Ond\v{r}ej Hul\'{i}k,}
\author[a]{Joris Raeymaekers}
\author[a]{and Orestis Vasilakis}
\affiliation[a]{Institute of Physics of the Czech Academy of Sciences, CEICO, \\ Na Slovance 2, 182 21 Prague 8, Czech Republic}
\affiliation[b]{Institute of Particle Physics and Nuclear Physics, Faculty of Mathematics and Physics, Charles University,  V Hole\v{s}ovi\v{c}k\'{a}ch 2, 180 00 Prague 8, Czech Republic}

\emailAdd{ondra.hulik@gmail.com}
\emailAdd{joris@fzu.cz}
\emailAdd{vasilakis@fzu.cz}

\abstract{
Motivated by the question of bulk localization in holography, 
we study the problem of constructing multi-centered solutions  in higher spin gravity which describe   point particles in the interior of AdS$_3$. In the Chern-Simons formulation these   take into account the backreaction after adding Wilson line sources. We focus on chiral solutions where only the left-moving sector is excited. In that case it is possible to choose a gauge where the dynamical variables are a set of Toda  fields living in the bulk. The problem then reduces to solving  the $\cala_{N-1}$ Toda equations with delta function sources, which in turn requires solving an associated  monodromy problem. We show that this  monodromy problem  is equivalent to the monodromy problem for a particular $\calw_N$ vacuum conformal block at large central charge. Therefore, knowledge of the $\calw_N$ vacuum block  determines the multi-centered solution. Our calculations go beyond the heavy-light approximation by including the backreaction of all higher spin particles.
} 
\arxivnumber{}
\preprint{}

\begin{document}

\maketitle

\section{Introduction}\label{secintro}

The question of how local bulk physics in anti-de-Sitter space emerges holographically from the boundary CFT  is an important one and has received considerable attention in recent years following the work  of HKKL \cite{Hamilton:2006az}.
Ultimately it has  ramifications for our understanding of the black hole interior and the information puzzle \cite{Almheiri:2012rt}\cite{Papadodimas:2012aq}. 

One of the basic  questions one can ask in this context is how a bulk state containing  a collection of classical particles, tracing out worldlines in the interior of Anti-de Sitter (AdS) space, is described in the dual CFT.  In the context of AdS$_3$/CFT$_2$, it was established \cite{Hartman:2013mia}\cite{Faulkner:2013yia}\cite{Fitzpatrick:2014vua}\cite{Hijano:2015rla}\cite{Hijano:2015zsa} that these are intimately linked to `classical' Virasoro conformal blocks at large central charge, in that the classical block computes the action of the classical particles.
This connection has been further explored and generalized in various ways.  In this work we will focus on the generalization to particles in higher spin AdS$_3$ gravity and their relation to classical blocks of the $\calw_N$ algebra \cite{Ammon:2013hba}\cite{Castro:2014mza}\cite{Besken:2016ooo}.

In most of the investigations into the relation between bulk particles and conformal blocks, a `heavy-light' approximation is made in which at most one of the particles is assumed heavy enough to backreact  on the geometry. To go beyond this approximation, one must consider fully backreacted multi-centered solutions in the bulk, which is a daunting task in general. In  \cite{Hulik:2016ifr}, it was shown that the backreaction problem in Lorentzian AdS$_3$ simplifies if one considers spinning particle trajectories which, on the boundary, only excite the left-moving sector. In this case, as we shall review in section \ref{secgrav} below, the spacetime contains a submanifold of constant negative curvature except at the particle locations where there are conical singularities. The problem then reduces to solving a Euclidean Liouville 
equation on the unit disk with ZZ boundary conditions \cite{Zamolodchikov:2001ah} and delta-function sources. This requires solving  a monodromy problem which is identical to the monodromy problem which determines the classical vacuum block in a specific channel. Therefore the multicentered solution  can in principle be constructed once the classical vacuum block is known, and vice versa. 

The goal of the current work is to generalize the above results to the
construction of multi-centered solutions in higher spin gravity theories,
coming from backreacting particles with higher spin charges.  This problem was  analyzed in \cite{Castro:2014mza}\cite{Besken:2016ooo} in
the heavy-light
limit. 
To  go beyond this approximation, we first recast the pure gravity results in Chern-Simons variables. We show that there exists a gauge which is  convenient for the problem at hand, in which the left-moving gauge field is a Lax connection for the Liouville  
theory. Similarly, for the higher spin case, we show in section \ref{secspin3} that we can go to a gauge where one of the gauge fields is a Lax connection for the $\cala_{N-1}$ Toda field theory.
One somewhat suprising feature to come out of this analysis is that the standard higher spin theory with $\textrm{sl}(N,\RR )$ gauge symmetry leads to a non-standard reality condition on the Euclidean Toda fields, while the 
standard   reality condition instead describes higher spin theory with $\textrm{su}(\lfloor \frac{N}{2} \rfloor , \lceil\frac{N}{2}\rceil)$ gauge symmetry.

The Toda fields that we  construct in this manner live in the bulk, and we  argue that by including suitable singular sources in the Toda equations we can take into account the backreaction of localized bulk particles.
Point particle sources can be coupled consistently to higher spin gravity in the form of Wilson lines  \cite{Witten:1989sx}\cite{Ammon:2013hba}, and we consider here  a class of `chiral'
particles which only couple to the left-moving gauge field.
The problem then reduces  to solving  the Toda system on the unit disk with delta function sources  under suitable boundary conditions. Upon taking into account the boundary conditions through a doubling trick, the problem can be reduced to a certain monodromy problem on the complex plane. We show in section \ref{secblocks} that the same monodromy problem arises in the determination of a classical $\calw_N$ vacuum block in a specific channel.
Therefore, from the knowledge of the classical $\calw_N$ vacuum block
we can in principle construct the multi-centered solution and vice versa.
While we treat the case of spin-3 gravity in the most detail, we discuss the generalization to the spin-$N$ case in section \ref{secspinN}.

To summarize our results in the context of bulk locality, we provide in this work a prescription to construct a state in the Lorentzian bulk theory containing localized particles from a classical block in a Euclidean  Toda CFT. 
This seems close in spirit to  H. Verlinde's  `CFT-AdS' idea, where states in AdS with localized particles or black holes are constructed directly from the 
 Euclidean  CFT  \cite{VerlindeCFTAdS} (see also \cite{Jackson:2014nla},\cite{Verlinde:2015qfa}). The present work can presumably be viewed as a concrete realization of this idea, albeit in a highly symmetric setting.

\section{Chiral Solutions in Pure Gravity}\label{secgrav}
In this section we review and  expand on an earlier observation \cite{Hulik:2016ifr} that a class of  `chiral' solutions in 2+1 dimensional AdS gravity can be conveniently described in terms of a  Liouville field living in the bulk. 
Here, the term chiral means  that, from the point of view of the boundary CFT, only the left-moving sector is excited. We will first derive this Liouville  description in the metric formulation, where it has a clear geometric origin,  and subsequently in Chern-Simons variables, which will facilitate the extension to higher spin gravity in the following sections.

\subsection{Chiral sector and bulk Liouville field} 
The general solution to pure AdS$_3$ gravity  which has a cylindrical boundary and satisfies Brown-Henneaux boundary conditions \cite{Brown:1986nw}, is parametrized by two arbitrary periodic functions $T(x_+), \bar T (x_-)$. The metric\footnote{Note that we work in units where the AdS radius is set to one.}
reads, in Fefferman-Graham coordinates,
\be \label{FGmetric}
ds^2 =  \frac{dy^2}{y^2} - \frac{1}{4 y^2} dx_+ dx_- + \frac{T}{ 4} dx_+^2 + \frac{\bar T}{ 4} dx_-^2 -  \frac{ y^2}{ 4} T \bar T dx_+ dx_- \,. 
\ee
Here, the boundary is at $y=0$, and $x_\pm = t \pm \f $ are light-cone coordinates on the boundary cylinder, with $(x_+, x_-) \sim ( x_+ + 2\p, x_- - 2\p)$.
For example, the global AdS$_3$ solution is given by
\be 
T = \bar T = -1.
\ee

In this work, we will be interested in the class of  `chiral' solutions, where $\bar T = -1$ takes the same\footnote{One can easily generalize this discussion to the case where  $\bar T$ is an arbitrary negative constant,
 as was done in \cite{Hulik:2016ifr}, though we will focus on $\bar T =-1$ in what follows.} value as for global AdS , while $T(x_+)$ can be  arbitrary. The metric (\ref{FGmetric}) can then be rewritten in the fibered form
\be 
ds^2 =- {1 \over 4} \left( dx_- + \half(y^{-2} - T y^2)  dx_+\right)^2 + {dy^2\over y^2} 
+ {1 \over 16}\left( y^{-2} + T y^2\right)^2 dx_+^2 \,.
\ee
The main observation is that the  metric on the 2D base manifold, given by the last two terms in the above expression, has constant negative curvature as one can easily verify. Therefore, we can make  a coordinate transformation which brings this base metric in conformal gauge, such that
\be 
 {dy^2\over y^2} 
+ {1 \over 16}\left( y^{-2} + T y^2\right)^2 dx_+^2 = e^{- 2 \f (z , \bar z)} dz d \bar z .\label{confgauge}
\ee
The field $\f (z, \bar  z)$ then satisfies the Liouville equation
\be 
\pa \bar \pa \f + e^{- 2 \f} =0, \label{Liouveq}
\ee 
and we will see below   that the function $T(x_+)$ is essentially the boundary value of the Liouville stress tensor (see (\ref{stressrelgrav})).
Note that the Liouville field $\f$ is a bulk field depending on the original coordinates $y$ and $x_+$.

For later convenience it will useful to work out this coordinate transformation  in more detail.  It follows from 
(\ref{confgauge}) that there exists a real function $\a(y,x_+)$ such that
 \bea 
 e^{-  \f } dz &=& e^{i \a}  \left( - {d y\over y}
 +{i\over 4}(y^{-2} + T y^2) dx_+ \right) \,. \label{coordtransf}
\eea
Furthermore, one can show that the 3D  Einstein equations imply  the relation
\be 
i( \pa_z \f dz - \pa_{\bar z} \f d \bar z) =   \half(y^{-2} - T y^2) d x_+ - d \a \,.
\label{alpharel}
 \ee
  In summary, for chiral solutions to pure 3D gravity the metric can be brought in the form
 \be 
 ds^2 = - {1 \over 4} \left( d\tilde t +  i ( \pa_z \f dz - \pa_{\bar z} \f d \bar z) \right)^2 + e^{- 2 \f (z , \bar z)} dz d \bar z\label{chiralmetr} \,,
 \ee 
 where $\f$ is a Liouville field satisfying (\ref{Liouveq}) and we have defined
 \be 
 \tilde t \equiv x_- + \a (y, x_+).
 \ee

\subsection{Chern-Simons variables}
 
 With a view towards generalizing this observation to higher spin gravity, it will be useful to describe the above change of variables  in the Chern-Simons formulation \cite{Achucarro:1987vz},\cite{Witten:1988hc}, where the gravitational field is described by a flat connection taking values in $\textrm{sl}(2, \RR) \oplus \overline{\textrm{sl}(2, \RR) }$. In this formulation the gauge connections $A, \bar A$ are related to the dreibein $e$ and the spin connection $\o$ as
 \be 
 e = A - \bar A, \qquad \o = A + \bar A.
 \ee
For solutions obeying the Chern-Simons equivalent of the Brown-Henneaux boundary conditions, the gauge connections can be taken to have the Fefferman-Graham form
\cite{Banados:1998gg}
\be 
A_{HW} = - L_0 {dy \over y} + \half \left( y^{-1} L_1 - T(x_+) y L_{-1} \right) dx_+, \qquad
\bar{A}_{HW} = \bar L_0 {dy \over y} - \half \left( y^{-1} \bar L_{-1} +  y \bar L_{1} \right) dx_-,\label{HWgrav}
\ee
where in the second formula we have already restricted to chiral solutions with $\bar T =-1$. Our claim is that we can perform a gauge transformation to a gauge where
the connections take the form:
\bea 
\tilde A &=& 2 e^{-\f } \mathfrak{Re} (dz) L_0-  \mathfrak{Im}\left( (\pa_z \f - e^{-\f} ) dz \right) L_1 
-  \mathfrak{Im}\left( (\pa_z \f + e^{-\f} ) dz \right) L_{-1}\label{Laxgrav1} \,, \\
\tilde{\bar A}&=& -\half  d \tilde t ( \bar L_1 +\bar L_{-1} ) .\label{Laxgrav}
\eea
The connection $\tilde A$ is a Lax connection for the Liouville  equation: one verifies that the  flatness  of 
$\tilde A$ is equivalent to the Liouville equation (\ref{Liouveq}). Indeed, the connection $\tilde A$ can be brought into the standard form in the literature for the Lax connection of Liouville theory. For this purpose we perform a  by a complex change of basis in the Lie algebra generated by \be V \equiv e^{{i \p\over 4} (L_1 - L_{-1})}, \label{Vdef}\ee 
resulting in the  the standard  Lax connection for the Liouville equation (see e.g. \cite{BabylonTalon})
\be 
A_T=V^{-1} \tilde A V =  ( \pa_z \f dz - \pa_{\bar z} \f d \bar z) L_0 - i  e^{-\f } \left( dz L_1 + d\bar z L_{-1}
\right)
.\label{Todagrav}
\ee
The right-moving connection becomes in this basis
\begin{equation}
  \bar{A}_T=  V^{-1}\tilde{\bar{A}}V=\im d\tilde{t}\bar{L}_0 \,.
\end{equation}
Here we have used that $V$ satisfies
$ V^{-1} \left( L_1 + L_{-1}\right) V =- 2 i L_0$. \\
The associated field strength  $ \tilde F = d\tilde A +\tilde  A\wedge \tilde  A$  satisfies 
\be 
F_T = V^{-1} \tilde F V =- 2 \left( \pa_z \pa_{\bar z} \f + e^{- 2 \f} \right) L_0 dz \wedge d \bar z .\label{Fgrav}
\ee
and its vanishing is indeed equivalent to the Liouville equation. Although the connections $A_T$ and $\bar{A}_T$ don't live in the original Lie algebra, they are useful to define since they give a field strength in the Cartan subalgebra and because they can be easily generalized for the case of arbitrary spin that we examine in section \ref{secspinN}.

It remains to describe the gauge transformation relating (\ref{HWgrav}) and (\ref{Laxgrav}) in more detail. A generic gauge transformation can be decomposed into a translation part involving the AdS translation generators $L_m - \bar L_m$, and a local Lorentz part 
involving the Lorentz generators $L_m + \bar L_m$. As shown in \cite{Witten:1988hc}, the translation part can be traded for  a coordinate transformation $(y, x_+, x_-) \to  (z, \bar z, \tilde t) $, which we take to be precisely the transformation derived before: 
\be 
z = z(y,x_+), \qquad \tilde t = x_- + \a(y,x_+)\label{coordgen} \,,
\ee
satisfying (\ref{coordtransf},\ref{alpharel}).
We then make an additional local Lorentz transformation with gauge parameter   $\L \bar \L$, where
\be
\L = e^{- \ln y L_0} e^{- {\a \over 2} ( L_1 + L_{-1})}, \qquad 
\bar \L = e^{-\ln y \bar L_0}e^{- {\a \over 2} (\bar L_1 + \bar L_{-1})}.\label{locallor}
\ee
One can  verify that this transformation takes  (\ref{HWgrav}) into (\ref{Laxgrav1},\ref{Laxgrav}), i.e. 
\be 
\tilde A = \L^{-1} (A_{HW} + d) \L, \qquad \tilde{\bar A} = \bar \L^{-1} (\bar{A}_{HW} + d) \bar \L .
\ee

To find the precise relation between the bulk Liouville field $\f(z,\bar z)$ and the boundary stress tensor $T(x_+)$, we will work out this gauge transformation near the boundary. First of all, by making a conformal transformation we may assume $z$ to take values in the unit disk, with the AdS boundary located at $|z|=1$. For global AdS, $T=-1$, this leads to
\be 
 \f = \ln(1- |z|^2) , \qquad \a=x_+ \label{liouvads}
\ee 
and the following relation between the coordinates:
\be 
z = {1-y^2\over 1+y^2} e^{ix_+}, \qquad \tilde t = x_-  + x_+.
\ee 
More generally, we allow Liouville solutions on the unit disk with the same blow-up behaviour near the boundary, i.e.
\be 
\f \sim \ln(1- |z|^2) + \calo(1).\label{ZZ}
\ee
This boundary condition (\ref{liouvads}) on the Liouville field was considered in \cite{Zamolodchikov:2001ah} and is often referred to as the `ZZ boundary condition'.  
In general, we can construct a holomorphic quantity from the Liouville field, the `Liouville stress tensor'
\be
\calt (z) = - 4 (\pa_z \f)^2 - 4 \pa_z^2 \f ,
\ee
where the overall normalization is chosen for later convenience.
Like $\f$, this is  a bulk quantity, though its value at the boundary turns out to be closely related to the boundary stress tensor  $T (x_+)$. Using the Liouville equation one shows that solutions which behave like (\ref{ZZ}) near $|z|=1$ can be expanded as
\be \label{ZZexpansion}
\f = \ln (1-|z|^2) - \left.\left( {z^2 \over 24} \calt (z) \right)\right \rvert_{|z|=1}(1-|z|^2)^2 + \calo (1-|z|^2)^3 .
\ee
Plugging this into eq.  (\ref{coordtransf}), one finds that the desired gauge transformation near the boundary is of the form (\ref{coordgen}), (\ref{locallor}) with 
\be 
z = \left(1-  2y^2 + 2 y^4 +  {2\over 3} \left( T(x_+) -2 \right) y^6 + \calo (y^8) )\right) e^{i x_+}, 
\qquad \a  = x_+  \label{coordnb}
\ee
and that the boundary stress tensor is related to the Liouville stress tensor as
\be 
T(x_+) =  e^{2 i x_+ } \calt  ( e^{ i x_+ } )-1.\label{stressrelgrav}
\ee
This relation was derived previously  in \cite{Hulik:2016ifr} using the metric formulation. It reflects  
the fact that, on the boundary, the coordinates $z$ and $x_+$ are related through $z=e^{ i x_+ }$: the relation (\ref{stressrelgrav}) is the standard conformal transformation of the stress tensor, with the  constant term arising from the Schwarzian derivative.

\subsection{Point-particle sources}
As motivated in the Introduction, in what follows we will not restrict ourselves to pure gravity, but we  will introduce point particle sources in the bulk which backreact on the metric. We want to choose the quantum numbers of the particles and their trajectories in such a way that  the backreacted  metric falls in the chiral class (\ref{chiralmetr}). The quantum numbers specifying the particle can be taken to be  taken to be the $\textrm{sl}(2, \RR) \oplus \overline{\textrm{sl}(2, \RR) }$  weights $(h, \bar h)$, or equivalently the particle mass $m= h + \bar h$ and spin $s = |h-\bar h|$. In order not to turn on the right-moving stress tensor $\bar T$, we take $\bar h=0$, in other words we consider `extremal' spinning particles with $m = s$.
We should keep in mind that the particle mass $m$ refers to the coefficient in front of the proper length
part of the action $m \int ds$. The extremal particles we are considering are in fact excitations\footnote{This seems  puzzling in the light that massless higher spin fields in AdS$_3$ do not have any local excitations, though one should keep in mind that they do have global boundary excitations. It would be interesting to elucidate  the particle limit of field theory in AdS$_3$.}   of massless  fields with spin,  since we recall that the field theory mass in AdS$_3$ is given by
\be 
M^2_{AdS} = (m-s)(m+s-2) .
\ee
Note that both concepts of mass  coincide in the limit  $m \gg s$.

 Spinning particles are described in metric variables by the Matthisson-Papapetrou-Dixon equations \cite{Mathisson:1937zz},\cite{Papapetrou:1951pa},\cite{Dixon:1970zza}. As was shown in \cite{Castro:2014tta}, in locally AdS$_3$ spaces spinning particles still move along geodesics. One can easily check that, in backgrounds  of the chiral  form  (\ref{chiralmetr}), the curves  of constant $z$ are geodesics, and these curves will be our spinning particle trajectories. In global AdS$_3$ these are helical trajectories spinning around the center of AdS$_3$ at constant radius.

 These considerations have a nice and elegant counterpart in Chern-Simons variables, which will easily generalize to the higher spin case. In the Chern-Simons description, point particles can be introduced by inserting Wilson lines $W_R (C) \bar W_{\bar R} (C)$ into the path integral \cite{Witten:1989sx}\cite{Ammon:2013hba}, with
 \be 
W_R (C)  = \tr_R \calp \exp \int_C A , \qquad \bar W_{\bar R} (C)  = \tr_{\bar R} \calp \exp \int_C \bar A  \,, \label{Wilsondef}
\ee
where $C$ is a curve and $R$ and $\bar R $ are (unitary, infinite-dimensional) $sl(2,\RR)$ representations with lowest weight $h$ and highest weight\footnote{We work in conventions where the AdS$_3$ translation generators are given by $P_m= L_m
- \bar L_m$. In particular, the energy is $L_0 - \bar L_0$, and the representations with energy bounded below are of the  (lowest weight, highest weight) type.} $- \bar h$ respectively. 
For the case  of interest where $\bar h=0$, the $\bar R$  representation is trivial and our spinning particles only couple to $A$. As advocated in \cite{Witten:1989sx}\cite{Ammon:2013hba}, it's useful to rewrite the trace over Hilbert space states in
(\ref{Wilsondef})  as a quantum mechanical path integral as we review
in Appendix \ref{WilsonParticles}. For large $h$, it's justified to make a saddle point approximation which amounts to adding a source term (\ref{leftaction}) to the classical Chern-Simons action.
As was shown in \cite{Castro:2014mza}, the resulting  description is indeed equivalent to the Matthisson-Papapetrou-Dixon equations.
As we show in Appendix \ref{WilsonParticles}, the effect of adding the spinning particles is to introduce delta-function sources on  the right hand side of the Liouville equation (\ref{Liouveq}):
\be
\pa \bar \pa \f + e^{-2\f } =  4\p G \sum_i m_i \d^2(z- z_i, \bar z- \bar z_i) \,, \label{Liouvsource}
\ee 
where the $m_i$ are the point-particle masses and $z_i$ their locations in the $z$ coordinate of (\ref{chiralmetr}). We see that the particles must be `heavy', in the
sense that $m_i \sim G^{-1} \sim c$ in order to produce backreaction. 

It was derived in \cite{Hulik:2016ifr} that solving the sourced Liouville equation with the boundary condition (\ref{ZZ}) amounts to solving a certain monodromy problem. It was also shown there that this monodromy problem is equivalent to the one which determines a certain CFT vacuum conformal block; knowledge of the conformal block therefore allows for the construction of the multi-centered solution and vice versa. In the following sections we will
generalize these results to the construction of multi-centered solutions in higher-spin gravity.

We end this section with some comments.
\begin{itemize}
 \item 
 We would like to emphasize that the multiparticle configurations we are considering here are specific to Lorentzian AdS$_3$. Indeed, our particle geodesics, which have constant
 AdS radial coordinate and constant value of $x_+$, do not analytically continue to geodesics in Euclidean AdS$_3$, except for the geodesic at the center of AdS$_3$.
 Unlike their Euclidean counterparts, our   Lorentzian geodesics do not reach the boundary and therefore do not correspond to localized sources in the boundary CFT; this is why their description as localized sources of the bulk Liouville field is particularly convenient.

 \item The setup described above might be viewed as a special case of Verlinde's `CFT-AdS'
 idea  \cite{VerlindeCFTAdS} (see also \cite{Jackson:2014nla},\cite{Verlinde:2015qfa}), where a state in the Lorentzian bulk theory is  directly corresponds to a state in the Euclidean CFT. In our case, the bulk state contains a number of spinning particles and corresponds to a state created by the insertion of heavy operators in Euclidean Liouville theory on the unit disk with ZZ boundary conditions. 
   Our bulk Liouville field $\f$ should also not be confused with the boundary Liouville field constructed in \cite{Coussaert:1995zp}, though the two are related:  the left-moving  components of their stress tensors obey the relation (\ref{stressrelgrav}).
\end{itemize}

\section{Chiral Solutions in Spin-3 Gravity}\label{secspin3}

Our main objective is to generalize the above results for pure gravity to the  Chern-Simons formulation of higher spin gravity. In this and the next section we will  perform the calculations explicitly for  the spin-3 theory, 
and we will extrapolate to arbitrary spin in  section \ref{secspinN}.

It is natural to guess that the role of Liouville theory in the pure gravity problem will be replaced by
Toda field theory in the higher spin case, and this will turn out to be so. One subtlety we will pay 
particular attention to is the following. One can consider two different higher spin theories based on the two 
noncompact real forms of the algebra $\cala_2$, namely  $\textrm{sl}(3,\RR)$ and $\textrm{su}(1,2)$. We will see that chiral solutions in these theories are described by  different   real sections of complex Toda field theory.  
Somewhat surprisingly,  we will see that the standard Euclidean Toda field theory
describes the non-standard higher spin theory with gauge algebra $\textrm{su}(1,2)$
while the
standard  $\textrm{sl}(3,\RR)$
higher spin theory requires a non-standard reality condition on the Toda fields.

\subsection{Spin-3 Gravity}
We start by recalling some standard facts about  gravity coupled to a massless spin-3 field  in  Lorentzian AdS$_3$, referring to 
\cite{Campoleoni:2010zq} for more details.
The action 
is given by
\be
S=S_{CS}[A]-S_{CS}[\bar A]~,
\ee
where
\be
S_{CS}[A]={k\over 4\pi}  \, \tr \int_{\M} \Big( A\wedge dA+ {2\over 3} A\wedge A\wedge A\Big)~.\label{CSaction}
\ee
where the gauge fields $A$ and $\bar A$ take values in a real form of the algebra $\cala_2$. There are two  real forms of this algebra which contain $\textrm{sl}(2,R)$ as a subalgebra and  give rise to a higher spin extension to gravity: these are the noncompact real forms $\textrm{sl}(3,\RR)$ and $\textrm{su}(1,2)$. The theory based on $\textrm{sl}(3,\RR)$ is the most prevalent in the literature as it leads \cite{Campoleoni:2010zq} to an asymptotic symmetry which is the standard Zamolodchikov $\calw_3$ algebra
\cite{Zamolodchikov:1985wn}, while the $\textrm{su}(1,2)$ leads to a different real form of the complex $\calw_3$ algebra which leads to problems with unitarity at the quantum level \cite{Campoleoni:2011hg}.

The two real forms
can be conveniently treated simultaneously by introducing a parameter $\s =\pm 1$, such that
\be
\s=1: \ \textrm{sl}(3,\RR) \oplus \overline{\textrm{sl}(3,\RR)}, \qquad \s=-1: \ \textrm{su}(1,2) \oplus \overline{ \textrm{su}(1,2)}.
\ee
The parameter $\s$ enters the  commutation relations which take the form:
\begin{align}
\, [ L_m, L_n ] =& (m-n) L_{m+n},&  m,n =& -1,0,1\\
\,  [ L_m, W_a ] =& (2 m- a ) W_{a+m},& a,b =& -2,\ldots ,2\\
\,   [ W_a, W_b ] =& -{\s \over 12} (a-b) (2 a^2+ 2 b^2 - a b -8)  L_{a+b} & &
\end{align}
and similarly for the barred generators. Note that the two real forms are related by the `Weyl unitary trick' of multiplying the spin-3 generators $W_a$ with a factor $\im$. 
An explicit matrix realization is given in Appendix \ref{matrea}. The connections $A$ and $\bar A$ are linear combinations of the generators $L_m, W_a$ (and their barred counterparts) with real coefficients. The generators $L_m$ form an $\textrm{sl}(2, \RR)$ subalgebra which is said to be `principally' embedded in the full algebra. The restriction of the gauge field to this subalgebra defines the gravitational subsector of the theory, while the coefficients of  $W_a$ and  $\bar W_a$  describe a massless spin-3 field. 

As shown in \cite{Campoleoni:2010zq}, field configurations obeying the higher spin equivalent of the  Brown-Henneaux boundary conditions \cite{Brown:1986nw} can be brought in a standard form which generalizes (\ref{HWgrav}) and is   called the `highest weight gauge'. Restricting once more to chiral solutions where $\bar A$ takes the same form as in global AdS$_3$, solutions   in this gauge  take the form
\bea 
A_{HW} &=& - {dy \over y} L_0 + \left( \half \left( y^{-1} L_1 -  T(x_+) y  L_{-1}\right) + W(x_+) y^2 W_{-2}\right) dx_+ \,, \\
\bar{A}_{HW} &=& \bar L_0 {dy \over y} - \half \left( y^{-1} \bar L_{-1} +  y \bar L_{1} \right) dx_-.\label{HWNis3}
\eea
Here, $T(x_+)$ and $W(x_+)$ are arbitrary periodic functions.

\subsection{The Toda field gauge}
We will now show that we can make a gauge transformation which brings the connection $A$ into the form of a Lax connection for $\cala_2$ Toda field theory, which will facilitate the construction of backreacted multi-particle solutions in the next subsection. More precisely, we will go to a gauge where
\begin{equation}\label{ATsl3}
\begin{split}
A_T= & V^{-1} \tilde A V =   {1 \over 4}\left( \pa_z\ (\f_1 + \f_2) dz - \pa_{\bar z}  (\f_1 + \f_2) d \bar z\right) L_0 \\
& - {\im \over 2 \sqrt{2}} \left( e^{{\f_1\over 2} -\f_2} + e^{{\f_2\over 2} -\f_1}\right)\left( dz L_1+  d\bar z L_{-1}\right)\\
& + {3 \over 4 \sqrt{\s}}\left( \pa_z\ (\f_1 - \f_2) dz - \pa_{\bar z}  (\f_1 - \f_2) d \bar z\right) W_0 \\
& + {\im \over \sqrt{2 \s} }  \left( e^{{\f_1\over 2} -\f_2} - e^{{\f_2\over 2} -\f_1}\right) \left(dz W_1+ d\bar z W_{-1}\right) \,,
\end{split}
\end{equation}
\begin{equation}
\bar{A}_T = V^{-1} \bar{\tilde A} V  = -\im d \tilde t \bar{L}_0 .\label{LaxNis3}
\end{equation}
As before, the only role of the complex change of basis generated by $V$ defined in (\ref{Vdef}) is to simplify the right hand side.
 The flatness condition $ \tilde F = d\tilde A +\tilde  A\wedge \tilde  A=0$  is equivalent to the $\cala_2$ Toda field equations
\begin{equation} \label{tod3}
    \begin{split}
        & \partial_z \partial_{\bar z} \phi_1 + e^{-2\phi_1+\phi_2}=0 \,, \\
        &  \partial_z \partial_{\bar z} \f_2+ e^{-2\phi_2+\phi_1}=0 \,.
    \end{split}
\end{equation}
If we initially view  $\f_1$ and $\f_2$ as complex-valued fields, the requirement that $\tilde A$ has real coefficients imposes a certain reality condition on the fields, which depends on the chosen real form, i.e. on the value of $\s$. On the components of $A_T$ in the expansion $A_T =\sum_m l_m L_m + \sum_a w_a W_a$, the condition that $\tilde A$  has real coefficients imposes that 
\be 
l_m^* = - l_{-m}, \qquad w_a^* =  w_{-a}.
\label{realrot}\ee 
Applying this to (\ref{LaxNis3}) one therefore finds that the appropriate reality condition is 
\begin{align}
\textrm{sl}(3,\RR),\  \s=& 1: &  \f_2 &= \bar \f_1 \nonu
\textrm{su}(1,2),\ \s=& -1: & \f_1, \f_2 &\in \RR.\label{realconds}
\end{align}
We observe that, somewhat surprisingly, the standard form of the Toda field equations, where $\f_1$ and $\f_2$ are real fields, is relevant for the nonstandard higher spin theory with algebra $\textrm{su}(1,2) \oplus \overline{ \textrm{su}(1,2)}$ and vice versa.

As shown in Appendix \ref{TodaTheory} \footnote{To simplify notation for this section we have relabeled the currents such that compared to Appendix \ref{TodaTheory} we have $\calt\equiv\mathcal{W}^{(2)}$,   $\calw\equiv\mathcal{W}^{(3)}$.} the Toda equations imply that the following combinations are purely holomorphic:
\bea \label{currentsTW}
\calt (z) &=& -(\pa \f_1 )^2 -(\pa \f_2 )^2 + \pa \f_1 \pa \f_2 - \pa^2 \f_1 - \pa^2 \f_2 \,, \\ 
\calw (z) &=& - (\pa \f_1 )^2 \pa \f_2 +  (\pa \f_2 )^2 \pa \f_1  -  \pa^2 \f_1  \pa \f_1 +  \pa^2 \f_2 \pa \f_2\nonu 
&&+ \half \left( - \pa^2 \f_1  \pa \f_2 +  \pa^2 \f_2  \pa \f_1 - \pa^3 \f_1 + \pa^3 \f_2 \right) \,.
\eea
One can show that $\calt$ transforms as a stress tensor under conformal transformations, while $\calw$ transforms as a spin-3 primary. 
We note from comparing (\ref{LaxNis3}) to the pure gravity expression (\ref{Laxgrav}) that the gravity subsector
is obtained by setting
\be 
\f_1 = \f_2 =  2 \f - \ln 2.
\ee
Note that we have $\calw =0$ in the gravity subsector as was to be expected.

From the pure gravity expression (\ref{liouvads}) and the Toda equations it follows  that the global AdS$_3$ solution is given by
\be 
\f_1 = \bar{\f}_2 = \ln {(1-|z|^2)^2\over 2} +n\frac{2\pi\im}{3}\, , \quad n\in \mathbb{Z}\label{AdSNis3}
\ee
and has $\calt = \calw=0$. For the $\s=-1$ case the fact that we have real Toda fields imposes that $n=0$. 
More generally, we allow Toda solutions on the unit disc with the same leading blow-up behaviour near the boundary circle.
These boundary conditions are the Toda  generalization of the ZZ boundary conditions  \cite{Zamolodchikov:2001ah}. Using the
Toda equations one shows that such solutions behave near the boundary as
\bea  \label{fexpansion}
\f_1 &=& \ln {(1-|z|^2)^2\over 2} + n\frac{2\pi\im}{3} - \left.\left( {z^2 \over 12} \calt \right)\right \rvert_{|z|=1} (1 -|z|^2)^2 - 
\left.\left( {z^2 \over 12} \calt  - {z^3 \over 60} \calw \right)\right \rvert_{|z|=1}  (1 -|z|^2)^3 + \ldots \nonu
\f_2 &=& \ln {(1-|z|^2)^2\over 2} - n\frac{2\pi\im}{3} - \left.\left( {z^2 \over 12} \calt \right)\right \rvert_{|z|=1} (1 -|z|^2)^2 - 
\left. \left( {z^2 \over 12} \calt  + {z^3 \over 60} \calw \right)\right \rvert_{|z|=1} (1 -|z|^2)^3 + \ldots \nonu \label{nbexpToda}
\eea 
Combining \eqref{fexpansion} with the reality conditions (\ref{realconds}) on the fields, we see that the currents  $\calt$ and $\calw $ should satisfy the following reality 
conditions on the boundary circle:
\begin{align}
\textrm{sl}(3,\RR),\  \s=& 1: & \left.\left( z^2  \calt \right)\right \rvert_{|z|=1} \in& \RR, & 
 \left.\left( z^3  \calw \right)\right \rvert_{|z|=1} \in& i\, \RR\nonu
\textrm{su}(1,2),\ \s=& -1: &\left.\left( z^2  \calt \right)\right \rvert_{|z|=1} \in& \RR, & 
 \left.\left( z^3  \calw \right)\right \rvert_{|z|=1} \in&  \RR.\label{realcurrents}
\end{align}

It remains to spell out the gauge transformation between the highest weight gauge (\ref{HWNis3}) and the Toda field gauge (\ref{LaxNis3}). We will only work out the near-boundary behaviour of this transformation, from which we will derive the precise relation between the
bulk Toda currents $\calt (z), \calw(z)$ and the boundary currents $T(x_+), W(x_+)$.
A generic gauge parameter contains a spin-2 and a spin-3 part, each of which  consists of  a `translation' and `local Lorentz' part. As in the gravity case, we will trade the spin-2 translation for a coordinate transformation
$(y, x_+, x_-) \to ( z, \bar z, \tilde t)  $. It turns out that, to the order required, the latter  coordinate transformation is unchanged from the pure gravity case, i.e. it still takes the form (\ref{coordnb}).
The required remaining gauge parameter turns out to be
\begin{equation}
\begin{split}
\L_3  =  \L \bar \L V &\exp \left( { W(x_+)  y^6} \left( {i \over 10} (e^{2 i x_+ } W_2 -e^{-2 i x_+ } W_{-2}) -
\right. \right.\\ & \left. \left.
-{3 \over 5}  (e^{ i x_+ } W_1 -e^{- i x_+ } W_{-1}) \right)  + \calo (y^8) \right) V^{-1} \,,
\end{split}
\end{equation}
where $\L, \bar \L$ are given in (\ref{locallor}). Performing the gauge transformation and making use of the near-boundary expansion (\ref{nbexpToda}) we find that (\ref{HWNis3}) transforms into  (\ref{LaxNis3}) with the currents related as
\be 
T(x_+) =e^{2 i x_+ } \calt  ( e^{ i x_+ } ) -1 , \qquad 
W(x_+) =  { i \over  \sqrt{\s} } e^{3 i x_+} \calw (e^{ix_+}).
\ee
As a check we note that $T(x_+)$ and  $W(x_+)$ are indeed real upon imposing the reality conditions (\ref{realconds}). These relations are consistent with the conformal transformation properties of the currents and the fact  that the boundary coordinates are related as $z = e^{ix_+}$.

\paragraph{Point-particle sources}
Having found convenient variables to describe chiral solutions in pure higher spin-3 gravity, we will now study solutions in the presence of  point-particle sources which are compatible with this chiral ansatz. Such higher-spin point particles are once  again described by  Wilson lines \cite{Ammon:2013hba}, and in order to preserve the chiral structure we study sources which couple only to $A$ and not to $\bar A$:
 \be 
W_R (C)  = \tr_R \calp \exp \int_C A \,. \label{Wilsonspin3}
\ee
 Such sources correspond to the spin-3 generalization of  extremal spinning particles.
 The representation $R$ is an infinite-dimensional  representation of $\textrm{sl}(2,\RR)$ resp. $\textrm{su}(1,2)$ built on a primary state $|h,w\rangle$ satisfying
 \be 
L_0 |h,w\rangle= h  |h,w\rangle , \qquad W_0 |h,w\rangle= w  |h,w\rangle, \qquad
L_{1} |h,w\rangle = W_{1,2} |h,w\rangle  =0 \,.
\ee
As shown in Appendix \ref{WilsonParticles}, in the presence of  such backreacting point particles the Toda equations acquire nontrivial delta function sources on the right hand side
\begin{equation}\label{spin3eoms}
\begin{split}
& \partial\bar{\partial}\phi_1 + \ex^{-2\phi_1 + \phi_2} = 16 \pi G\sum_i\alpha_1^{(i)}\delta^2\left(z-z_i,\bar{z}-\bar{z}_i\right) \,, \\
& \partial\bar{\partial}\phi_2 + \ex^{-2\phi_2 + \phi_1} = 16\pi G\sum_i\alpha_2^{(i)}\delta^2\left(z-z_i,\bar{z}-\bar{z}_i\right) \,,
\end{split}
\end{equation}
where the parameters $\a^{(i)}_1, \a^{(i)}_2$ are related to the conformal dimension and spin-3 charge of the $i$-th particle as
\begin{equation}\label{spin3constraints}
\begin{split}
& (\a_1^{(i)})^2 - \a_1^{(i)} \a^{(i)}_2 + (\a_2^{(i)})^2 =\frac{1}{4}\left( (h^{(i)})^2 + 3 \s (w^{(i)})^2\right)  \,, \\
& \left(\alpha^{(i)}_2-\alpha^{(i)}_1\right)\alpha^{(i)}_1\alpha^{(i)}_2 = { \im \over \sqrt{4\s}}  w^{(i)}\left((h^{(i)})^2 -\s  (w^{(i)})^2\right) \,.
\end{split}
\end{equation}
It's a useful consistency check that these sources are compatible with the reality conditions (\ref{realconds})
which  for $\s=1$  require that $\a^{(i)}_2 = \bar{\a}^{(i)}_1$, and   for $\s=-1$ that both $\a^{(i)}_1$ and $\a^{(i)}_2$ are real. The  equations (\ref{spin3constraints}) for
 $\a^{(i)}_{1,2}$  are indeed compatible with these reality conditions.

\subsection{Properties of the solutions}

\paragraph{The associate ODE problem:} 
Toda field theory is an integrable system and its solutions can be constructed from solutions to a set of
auxiliary ordinary differential equations (ODEs).
For more details on these auxiliary ODEs we refer the reader to Appendix \ref{TodaTheory}. Here we give just the essentials needed to make the discussion  coherent.

The  $\mathcal{A}_2$-Toda field theory has two systems of auxiliary ODEs associated with it. One shows that the fields $\ex^{\phi_i}$ satisfy the following sets of equations 
\begin{align}
\left[ \partial^3 + \calt  \partial + \calw + \half \pa \calt \right] \ex^{\phi_1}& =0, & \left[\bar \partial^3 + \bar \calt  \bar \partial + \s \bar \calw + \half \bar \pa \bar \calt \right] \ex^{\phi_1}& =0,\label{ODE1}\\
\left[ \partial^3 + \calt  \partial - \calw + \half \pa \calt \right] \ex^{\phi_2}& =0, & \left[\bar \partial^3 + \bar \calt  \bar \partial - \s \bar \calw + \half \bar \pa \bar \calt \right] \ex^{\phi_2}& =0,\label{ODE2}
\end{align}
with $\calt(z) $  the stress tensor  and $\calw (z)$ the primary spin-3 current  given in (\ref{currentsTW}). We will denote by $\psi_i(z),\ i = 1,2,3$  a set of linearly independent solutions of the equation
\be \label{psieq}
\left[ \partial^3 + \calt  \partial + \calw + \half \pa \calt \right] \psi(z)=0
\ee
and by $\chi^i(z),\ i = 1,2,3$  a set of  independent solutions of 
\be  \label{chieq}
\left[ \partial^3 + \calt  \partial - \calw + \half \pa \calt \right] \chi(z)=0 \,.
\ee
We will now show that we can build Toda solutions $e^{\f_i}$ from the $\psi_i$, $\chi^i $ in a holomorphically  factorized form.

\paragraph{Solutions for real Toda fields -- ${\rm su}(1,2)$ case: }
From \eqref{ODE1} with $\s=1$ we have
\begin{equation}\label{phi1bilinear}
    \ex^{\phi_1} =\Psi^{\dagger}\Lambda\Psi \,,
\end{equation}
where $\Psi = \left(\psi_1,\psi_2,\psi_3\right)^T$ and 
\begin{equation}\label{sl3Lambda}
    \Lambda = \textrm{diag}\left(1,-1,1\right)
\end{equation}
as specified in appendix \ref{TodaTheory}. Similarly from \eqref{ODE2} with $\s=1$
we deduce
\begin{equation}\label{phi2bilinear}
    \ex^{\phi_2} = X^{\dagger}\Lambda_2 X,
\end{equation}
where $X = \left(\chi_1,\chi_2,\chi_3\right)^T$.
By substituting \eqref{phi1bilinear} into the first Toda equation \eqref{tod3} and solving for $\ex^{\phi_2}$ we find that $\Lambda_2 = \Lambda$ and 
\begin{equation}\label{chiminorpsi}
    \chi_a = \epsilon_{abc}\Lambda^{bb'}\Lambda^{cc'}  \psi_{b'}\, , \partial\psi_{c'}
\end{equation}
while by substituting \eqref{phi1bilinear}, \eqref{phi2bilinear}, \eqref{chiminorpsi} in the second Toda equation \eqref{tod3} we independently derive that
\begin{equation}
    \textrm{det}\Lambda  = -1\, ,
\end{equation}
which is consistent with our choice \eqref{sl3Lambda}.

\paragraph{Solutions for complex conjugate Toda fields -- ${\rm sl}(3,\mathbb{R})$ case: }
From \eqref{ODE1} and \eqref{ODE2} with $\s=-1$ we have
\begin{equation}\label{phisl3Rbilinear}
    \ex^{\phi_1} = \psi_i \bar{\chi}^j \,,
    \qquad
    \ex^{\phi_2} = \bar{\psi}_i \chi^i \,.
\end{equation}
By substituting \eqref{phisl3Rbilinear} into the first Toda equation \eqref{tod3} and solving for $\ex^{\phi_2}$ we find that
\begin{equation}
\chi^a = -\epsilon^{abc} \psi_{b} \partial \psi_{c}\, .
\end{equation}
Thus for the Toda fields we have
\begin{equation}\label{sl3Todapsi}
    \ex^{\phi_1} = \ex^{\bar{\phi}_2} = -\epsilon^{abc}\psi_a\bar{\psi}_b\bar{\partial}\bar{\psi}_c \,.
\end{equation}
%

\subsection{Properties of the currents}\label{sl3currentsproperties}
\paragraph{Pole structure:} If near the sources $z_i$ we can neglect the potential term compared to the kinetic term, then the Toda fields $\phi_i$ behave near $z=z_i$ as 
\begin{equation}\label{Todanearsource}
\phi_j\sim 16G\alpha_j^{(i)}\ln|z-z_i| \,.
\end{equation}
Thus the solution of the Toda equations on the disk specifies meromorphic (quasi-)primary currents $\mathcal{T}$ and $\mathcal{W}$. The sources $\alpha^{(i)}_j$ fix the most singular terms in $\calt, \calw$ which take the form
\begin{equation}\label{TWmostsingular}
    \mathcal{T}(z) =\sum_{i=1}^K\frac{\epsilon_i^{(\mathcal{T})}}{(z-z_i)^2}+ \ldots \, , \quad \mathcal{W}(z) =\sum_{i=1}^K\frac{\epsilon_i^{(\mathcal{W})}}{(z-z_i)^3}+ \ldots
\end{equation}
where the ellipses denote lower order poles and a regular part. The constants $\epsilon_i^{\mathcal{T}}$ and $\epsilon_i^{\mathcal{W}}$ are specified in terms of $\alpha_1^{(i)}$ and $\alpha_2^{(i)}$ by examining the form of the Toda fields $\phi_1$ and $\phi_2$ near a specific source.
\begin{equation}
    \begin{split}
        & \epsilon_i^{(\mathcal{T})} =8G\left(a_1^{(i)}+a_2^{(i)}\right) + 64G^2\left(a_1^{(i)}a_2^{(i)} - (a_1^{(i)})^2 - (a_2^{(i)})^2  \right)  \,, \\
        & \epsilon_i^{(\mathcal{W})} = 8G\left(a_2^{(i)}-a_1^{(i)}\right)\left(1-8Ga_1^{(i)}\right)\left(1-8Ga_2^{(i)}\right) \,.
    \end{split}
\end{equation}
\paragraph{Doubling trick:} 
Having established the pole structure of the currents $\calt (z)$ and $\calw (z)$ as functions on the unit disk, we should also make sure that they obey the reality conditions
(\ref{realcurrents}) on the unit circle. As usual, this is imposed by using a `doubling trick' and extending $\calt (z)$ and $\calw (z)$  to meromorphic functions on the complex plane in a suitable manner.  Using the Schwarz reflection principle we find  that the appropriate reflection  properties are
\begin{equation}\label{sl3currentreflection}
    \mathcal{T} (z)  = \frac{1}{z^4}\overline{\mathcal{T}}\left(\frac{1}{z}\right) \,, \quad \mathcal{W}(z) = -\frac{\s}{z^6}\overline{\mathcal{W}}\left(\frac{1}{z}\right) \,.
\end{equation}
In particular, this means that the currents  have poles both at the $z_i$ and at their image points $\bar{z}_i^{-1}$. Thus they are of the form
\begin{equation}\label{W2poles}
\mathcal{T}= \sum_{i=1}^K\left(\frac{\epsilon^{(\mathcal{T})}_i}{\left(z-z_i\right)^2} + \frac{\tilde{\epsilon}^{(\mathcal{T})}_i}{\left(z-1/\bar{z}_i\right)^2} + \frac{c_i^{(\mathcal{T},1)}}{z-z_i} + \frac{\tilde{c}_i^{(\mathcal{T},1)}}{z-1/\bar{z}_i} \right) \,,
\end{equation}
\begin{equation}\label{W3poles}
\mathcal{W} = \sum_{i=1}^K\left(\frac{\epsilon^{(\mathcal{W})}_i}{\left(z-z_i\right)^3} + \frac{\tilde{\epsilon}^{(\mathcal{W})}_i}{\left(z-1/\bar{z}_i\right)^3} + \frac{c_i^{(\mathcal{W},2)}}{\left(z-z_i\right)^2} + \frac{\tilde{c}_i^{(\mathcal{W},2)}}{\left(z-1/\bar{z}_i\right)^2} + \frac{c_i^{(\mathcal{W},1)}}{z-z_i} + \frac{\tilde{c}_i^{(\mathcal{W},1)}}{z-1/\bar{z}_i} \right) \,,
\end{equation}
where in the above equations we assumed for simplicity that there is no source located at the origin and as a result there is also no image source at infinity. 

The parameters $c_i^{(s,l)}, \tilde{c}_i^{(s,l)}$ are called accessory parameters and are not directly determined by the $\alpha^{(i)}_j$. Instead we will see below that they are determined by solving a monodromy problem which arises from demanding that the Toda fields are single-valued. 
Not all the parameters in (\ref{W2poles}) and (\ref{W3poles}) are independent however, since we still have to impose the reflection properties (\ref{sl3currentreflection}). 
We will now determine the number of independent  accessory parameters after imposing the 
reflection properties.
Substituting (\ref{W2poles}) and (\ref{W3poles}) into (\ref{sl3currentreflection}) and requiring equality of the poles at each order near $z_i$ we get from $\mathcal{T}$
\begin{equation}\label{W2poleequality}
\epsilon^{(\mathcal{T})}_i-\bar{\tilde{\epsilon}}^{(\mathcal{T})}_i =0 \, , \quad 2\epsilon^{(\mathcal{T})}_i+c_i^{(\mathcal{T},1)}z_i + \frac{\bar{\tilde{c}}^{(\mathcal{T},1)}_i}{z_i}=0
\end{equation}
and from $\mathcal{W}$
\begin{equation}\label{W3poleequality}
\begin{split}
& \epsilon^{(\mathcal{W})}_i- \s \bar{\tilde{\epsilon}}^{(\mathcal{W})}_i =0 \,, \\
& 3\epsilon^{(\mathcal{W})}_i+c_i^{(\mathcal{W},2)}z_i + \s \frac{\bar{\tilde{c}}^{(\mathcal{W},2)}_i}{z_i}=0 \,, \\
& 6\epsilon^{(\mathcal{W})}_i +4\bar{{c}}^{(\mathcal{W},2)}_i z_i + c^{(\mathcal{W},1)}_iz_i^2-\s \frac{\bar{\tilde{c}}^{(\mathcal{W},1)}_i}{z_i^2} =0 \,.
\end{split}
\end{equation}
By substituting (\ref{W2poleequality}) and (\ref{W3poleequality}) into (\ref{W2poles}) and (\ref{W3poles}) respectively and requiring (\ref{sl3currentreflection}) to hold at the origin we get for $\mathcal{T}$
\begin{equation}\label{W2originreg}
\begin{split}
& \sum_{i=1}^K\left(c^{(\mathcal{T},1)}_i + \tilde{c}^{(\mathcal{T},1)}_i \right)=0 \,, \\
& \sum_{i=1}^K\mathfrak{Im}\left(\e_i + c^{(\mathcal{T},1)}_iz_i\right) = 0
\end{split}
\end{equation}
and from $\mathcal{W}$
\begin{equation}\label{W3originreg}
\begin{split}
& \sum_{i=1}^K\left(c^{(\mathcal{W},1)}_i + \tilde{c}^{(\mathcal{W},1)}_i \right)=0 \,, \\
& \sum_{i=1}^K\left(c_i^{(\mathcal{W},2)} + \tilde{c}^{(\mathcal{W},2)}_i + c^{(\mathcal{W},1)}_iz_i+\frac{\tilde{c}^{(\mathcal{W},1)}_i}{\bar{z}_i}\right) = 0 \,, \\
& \sum_{i=1}^K\mathfrak{Re}\left(\sqrt{\s} ( \epsilon^{(\mathcal{W})}_i + 2 c_i^{(\mathcal{W},2)}  z_i  + c_i^{(\mathcal{W},1)} z_i^2) \right) = 0 \,.
\end{split}
\end{equation}
One can check that these conditions guarantee that (\ref{sl3currentreflection}) holds everywhere. 
Regularity at infinity requires that $\calt$ falls of as $z^{-4}$ and $\calw$  as $z^{-6}$, which is implied by the reflection conditions (\ref{sl3currentreflection}) and the fact that  $\calt, \calw$ are regular in the origin. So this does not lead to any further conditions.

After taking into account these constraints we are left with $2K-3$ independent real accessory parameters coming from $\mathcal{T}$ and $4K-5$ parameters coming from $\mathcal{W}$, giving in total $6K-8$ real accessory parameters to be specified by solving the monodromy problem. Note that the above conditions were derived in the assumption that none of the poles are in the origin. It  is straightforward to generalize to the case of a pole in the origin and to check that the number of independent accessory parameters remains unchanged.

\subsection{Doubling trick for $\psi_i$}

The reflection property of the currents \eqref{sl3currentreflection}, together with the associate ODEs, implies there should be a reflection property for the $\psi_i$. \\
For the $\textrm{su}(1,2)$ case we find
\begin{equation}\label{psireflection}
    \psi_a(z) = - z^2\epsilon_{abc}\Lambda^{b \tilde{b}} \Lambda^{c \tilde{c}} \bar{\psi}_{\tilde{b}} (1/z) \partial_{\frac{1}{z}}\bar{\psi}_{\tilde{c}} (1/z)
\end{equation}
while for the $\textrm{sl}(3,\mathbb{R})$ case we have
\begin{equation}\label{sl3Rpsireflection}
    \psi_a(z) = \im \ex^{-\im\frac{\pi}{6}} z^2\bar{\psi}_a\left(1/z\right) \,.
\end{equation}
To prove these relations we  start from  the associated ODE (\ref{psieq}) satisfied by $\psi_i(z)$:
%
\begin{equation}\label{ODEchiprimary}
   \left( \partial^3_z + \mathcal{T}(z)\partial_z + \mathcal{W}(z)+ \frac{1}{2}\partial_z\mathcal{T}(z)\right)\psi_i(z) = 0 \,.
\end{equation}
Under a conformal transformation $z \to \tilde z = f(z)$ the fields transform as
\be 
\tilde \calt (\tilde z) = (f')^{-2} \left( \calt (z) - 2 S(f,z) \right), \qquad 
\tilde \calw (\tilde z) = (f')^{-3}  \calw (z), \qquad \tilde \psi_i (\tilde z) = f' \psi_i (z) .\ee
The last transformation can be derived from the fact that $\psi_i$ should transform like $e^{\f_1}$, i.e. as a primary of weight $-1$, under holomorphic reparametrizations in order for (\ref{ODEchiprimary}) to be conformally invariant.
Using these transformations, we perform a conformal transformation with $f(z) = 1/z$ on 
(\ref{ODEchiprimary}), and use the reflection conditions (\ref{sl3currentreflection})
to obtain, after a relabeling of the coordinate:
\be \label{psieqrefl}
 \left( \partial^3_{\bar z} + \bar \calt (\bar z)\partial_{\bar z} + \s \bar \calw (\bar z)+ \frac{1}{2}\partial_{\bar z} \bar\calt (\bar z)\right)\left( \bar z^2 \psi_i\left({1\over \bar z}\right)\right) =0 \,.
 \ee
For the $\textrm{su}(1,2)$ case, upon choosing $\sigma=- 1$ 
and comparing with the ODE (\ref{chieq}) satisfied by $\chi^i$, we see that there should exist a constant matrix $S$ such that
\begin{equation}\label{psichirelation}
    \Psi(z) = Sz^2\bar{X}\left(\frac{1}{z}\right).
\end{equation}
Here,  $S$ is a proportionality matrix whose exact form is independent of the particular solution, and can be fixed by examining the simple solution displayed in appendix \ref{simplesolution}. For the case under consideration we find
\begin{equation}
    S=- \mathbb{I} 
\end{equation}
which upon substituting into \eqref{psichirelation} leads to \eqref{psireflection}.\\
Similarly, for the $\textrm{sl}(3,\mathbb{R})$ case, upon setting $\sigma= 1$
in (\ref{psieqrefl}) we see that there should exist a constant matrix $\tilde S$ such that
\begin{equation}\label{psipsirelation}
    \Psi(z) = \tilde{S}z^2\bar{\Psi}\left(\frac{1}{z}\right) \,.
\end{equation}
Once again, the form of $\tilde S$ is fixed by examining the simple solution displayed in appendix \ref{simplesolution}. For the case under consideration we find
\begin{equation}
    \tilde{S}=\im \ex^{-\im\frac{\pi}{6}}\mathbb{I}
\end{equation}
which upon substituting into \eqref{psipsirelation} leads to \eqref{sl3Rpsireflection}.\\

\subsection{The monodromy problem}\label{subsectionsl3monodromy}

\subsubsection{Single-valuedness of the Toda fields}

Since the currents are meromorphic functions, the associated ODE contains singularities. Thus after encircling a singular point $z_i$ the solution transforms as 
\begin{equation}
    \Psi \rightarrow M_i\Psi \,,
\end{equation}
where $M_i\in \textrm{SL}(3,\mathbb{C})$ is a monodromy matrix. However further restrictions are imposed on this monodromy matrix by requiring that the Toda fields are single valued. We will see that in each case the monodromy matrix must be an element of one of the real sections of $\textrm{SL}(3,\mathbb{C})$.
\paragraph{Real Toda fields -- $\textrm{su}(1,2)$ case:}
As we derived in \eqref{phi1bilinear} we have
\begin{equation}
    \ex^{\phi_1} = \Psi^{\dagger}\Lambda\Psi \,.
\end{equation}
Thus for $\ex^{\phi_1}$ to be single-valued we should demand that
\begin{equation}
    M_i^{\dagger}\Lambda M_i = \Lambda
\end{equation}
with $\Lambda = \textrm{diag}\left(1,-1,1\right)$. This means that we must have $M_i\in \textrm{SU}(1,2)$. In other words in order for the solution of the Toda system to be single-valued we must adjust the accessory parameters such that all the monodromy matrices $M_i$ are elements of $\textrm{SU}(1,2)$.

\paragraph{Complex conjugate Toda fields --  ${\rm sl}(3,\mathbb{R})$ case:}  

For the first field in the Toda system we have \eqref{sl3Todapsi}
\begin{equation}
\ex^{\phi_1} = -\epsilon^{abc} \psi_a \bar{\psi}_b \bar{\partial} \bar{\psi}_c \, .
\end{equation}
In order for it to be single-valued we should have that
\begin{equation}
\epsilon^{abc} {M_a}^d {\overline{M}_b}^e {\overline{M}_c}^f =\epsilon^{def} \, .
\end{equation}
which can be rewritten as
\begin{equation}\label{sl3monocondition}
    \frac{1}{2!}\epsilon_{gef}\epsilon^{abc} {M_a}^d {\overline{M}_b}^e {\overline{M}_c}^f=\delta^d_g \,.
\end{equation}
At this point we make use of the definition of the adjugate matrix
\begin{equation}\label{sl3adjdef}
   (\textrm{adj}{F)^b}_a = \frac{1}{2!}\epsilon^{bc_1c_2}\epsilon_{ad_1d_2}{F^{d_1}}_{c_1}{F^{d_2}}_{c_2} \,.
\end{equation}
In the present case we have $F^T=\overline{M}$ and because $M\in\textrm{SL}(3,\mathbb{C})$ we have
\begin{equation}
    \textrm{adj}M = M^{-1} \,.
\end{equation}
Thus \eqref{sl3monocondition} becomes
\begin{equation}
    M=\overline{M} \,.
\end{equation}
Which means that $M_i\in \textrm{SL}(3,\mathbb{R})$..

\subsubsection{Monodromies of image points}\label{imagemon}

Here we will prove that the monodromy matrix of a contour that encircles a singular point $z_i$ and its image $1/\bar{z}_i$ (and none of the other singularities) is the identity matrix. This property will be important in making the connection to classical $\calw_3$ blocks in section \ref{secblocks}. To do so lets us first consider the monodromy matrix $M_{\gamma_i}$ of a point $z_i$ which is encircled by a counterclockwise contour $\gamma_i$ which has a base point $p$ at the boundary. The mirror contour $\bar{\gamma}_i$ encircles the image point $1/\bar{z}_i$ in an opposite, clockwise, orientation  and results in a monodromy matrix $M_{\bar{\gamma}_i}$. We denote the contour encircling the image point counterclockwise as $\bar{\gamma}_i^{-1}$. Then
\begin{equation}\label{mirrormonorelation}
    M_{\bar{\gamma}_i^{-1}}=M^{-1}_{\bar{\gamma}_i} \,.
\end{equation}
Thus we need to show that
\begin{equation}
    M_{\gamma_i}M_{\bar{\gamma}_i^{-1}} = 1
\end{equation}
which due to \eqref{mirrormonorelation} is equivalent to showing that
\begin{equation}\label{origmirrormonorelation}
    M_{\gamma_i} = M_{\bar{\gamma}_i} \,.
\end{equation}
We will use the reflection property of $\psi_i$ to show that this is indeed the case. For simplicity we are going to drop from $M$ the subscript $\gamma_i$ and adopt a new notation such that
\begin{equation}
    M_{\gamma_i} \equiv M \, , \quad  M_{\bar{\gamma}_i}\equiv N \,.
\end{equation}

\paragraph{Real Toda fields -- $\textrm{su}(1,2)$ case:}
Let us first rewrite \eqref{psireflection} as 
\begin{equation}\label{psireflection2}
    \psi_a(z) = - z^2\epsilon_{abc}\Lambda^{b \tilde{b}} \Lambda^{c \tilde{c}} \overline{\psi_{\tilde{b}} (1/\bar{z})}\, \overline{\partial_{\frac{1}{\bar{z}}}}\,\overline{\psi_{\tilde{c}} (1/\bar{z})} \,.
\end{equation}
Then after encircling the point $z_i$ we have 
\begin{equation}\label{mirrormonopsi1}
    {M_a}^f\psi_f = -z^2\epsilon_{ab_1b_2}\Lambda^{b_1c_1}\Lambda^{b_2c_2}{\overline{N}_{c_1}}^{d_1}{\overline{N}_{c_2}}^{d_2}\Lambda_{d_1e_1}\Lambda_{d_2e_2}\overline{\psi}^{e_1}\overline{\partial\psi}^{e_2} \,.
\end{equation}
At this point we make use of the definition of the adjugate matrix \eqref{sl3adjdef}
which we can rewrite as
\begin{equation}\label{sl3adj2}
    \epsilon_{bf_1f_2}(\textrm{adj}{F)^b}_a = \epsilon_{ad_1d_2}{F^{d_1}}_{f_1}{F^{d_2}}_{f_2} \,.
\end{equation}
In the present case $F=\Lambda\overline{N}\Lambda$. Thus \eqref{mirrormonopsi1} becomes
\begin{equation}\label{mirrormonopsi2}
    {M_a}^f\psi_f = -z^2\epsilon_{be_1e_2}{(\textrm{adj}(\Lambda\overline{N}\Lambda))^b}_a\overline{\psi}^{e_1}\overline{\partial\psi}^{e_2} \,,
\end{equation}
which because of \eqref{psireflection2} gives
\begin{equation}\label{mirrormonopsi3}
    {M_a}^f\psi_f = {(\textrm{adj}(\Lambda\overline{N}\Lambda))^b}_a \psi_b \,.
\end{equation}
Now because $N\in \textrm{SU}(1,2)$ we have 
\begin{equation}
    \Lambda N^{\dagger}\Lambda = N^{-1}
\end{equation}
and 
\begin{equation}
    \textrm{adj}N = N^{-1}
\end{equation}
since $\textrm{det}N=1$. Using in addition that the transpose of an adjugate matrix is the adjugate of the transpose we have for the right hand side of \eqref{mirrormonopsi3}
\begin{equation}
    {(\textrm{adj}(\Lambda\overline{N}\Lambda))^b}_a = {(\textrm{adj}(\Lambda N^{\dagger}\Lambda))_a}^b = {(\textrm{adj}N^{-1})_a}^b = {N_a}^b \,.
\end{equation}
Substituting back in \eqref{mirrormonopsi3} we indeed get \eqref{origmirrormonorelation}.

\paragraph{Complex conjugate Toda fields --  ${\rm sl}(3,\mathbb{R})$ case:} 
Again we start by rewriting \eqref{sl3Rpsireflection} as
\begin{equation}\label{barsl3Rpsireflection}
    \psi_a(z) = \im \ex^{-\im\frac{\pi}{6}} z^2\overline{\psi_a\left(1/\bar{z}\right)} \,.
\end{equation}
Then after encircling the point $z_i$ we have
\begin{equation}
    {M_a}^b \psi_b = \im \ex^{-\im\frac{\pi}{6}} z^2 {\overline{N}_a}^c \overline{\psi_c \left(1/\bar{z}\right)} \,.
\end{equation}
Since $N$ is real, by combining the last equation with \eqref{barsl3Rpsireflection},  we immediately see that \eqref{origmirrormonorelation} holds.

\subsubsection{Monodromy around all points}
In this section we will show that the monodromy of a contour encircling all points within the unit disk takes values in one of the real forms $\textrm{SL}(3,\mathbb{R})$ resp. $\textrm{SU}(1,2)$, without making the assumption that the monodromy around each of the points belongs to a real form of $\textrm{SL}(3,\mathbb{C})$. This fact will be of use in the counting of accessory parameters that follows this section.

Let us denote by $\Gamma$ a counterclockwise contour encircling all the points within the unit disk, which has a base point $p$ at the boundary. The mirror contour, $\bar{\Gamma}$ encircles all the image points in an opposite, clockwise, orientation. We denote the contour encircling all the image points counterclockwise as $\bar{\Gamma}^{-1}$. Then
\begin{equation}\label{mirrormonorelationall}
    M_{\bar{\Gamma}^{-1}}=M^{-1}_{\bar{\Gamma}} \,.
\end{equation}
For a counterclockwise contour encircling all of the points and their images we have
\begin{equation}
    M_{ \text{All}} = M_{\Gamma}M_{\bar{\Gamma}^{-1}}=1
\end{equation}
which because of \eqref{mirrormonorelationall} implies
\begin{equation}\label{origmirrormonorelationall}
    M_{\Gamma} = M_{\bar{\Gamma}} \,.
\end{equation}
The latter is similar to \eqref{origmirrormonorelation}. Then the rest of the proof follows the same steps in the case of monodromies of image points by inverting the logic of that section. Specifically, instead of starting with the group the monodromy matrix belongs to and trying to prove \eqref{origmirrormonorelation}, we start with \eqref{origmirrormonorelationall} and we find the conditions  that specify the group the monodromy matrix belongs to.

\subsubsection{Parameter counting and monodromy reduction} \label{parcunt}
We will now show that the condition that the monodromy matrices take values in 
$\textrm{SL}\left(3,\mathbb{R}\right)$ resp. $\textrm{SU}(1,2)$, 
in the generic situation, imposes precisely as many constraints as the number of available undetermined accessory parameters in the ODE, namely  $6K-8$. Therefore, for generic values of the particle positions and quantum numbers, we expect our equations to have a unique solution.  
 To count the number of required constraints, we follow the following logic (see \cite{Hulik:2016ifr}): we first compute the dimension of the space in which the monodromy matrices around the $K$ singular points in the unit disk take values if the accessory parameters are arbitrary, and subtract from this the  dimension of the space in which they take values when the $\textrm{SL}\left(3,\mathbb{R}\right)$
(resp.  $\textrm{SU}(1,2)$) condition is imposed. 

To compute the dimension of the former space, we note that the  monodromy matrix
$M_i$ takes values in $\textrm{SL}\left(3,\mathbb{C}\right)$ which has real dimension  16, 
leading to $16 K$ parameters. Not all of these are independent however, since the coefficients  $\epsilon^{(s)}_i$ of the most singular pole terms in $\calt$ and $\calw$ are fixed. The latter fix the trace invariants $\tr M_i$ and $\tr M_i^2$  and 
therefore subtract $4K$ parameters. Furthermore, as we proved, the reflection property for $\psi_i$ implies that the monodromy, when encircling all singular points in the unit disk, must take values in one of the real forms $\textrm{SL}\left(3,\mathbb{R}\right)$ resp. $\textrm{SU}(1,2)$ 
, thus subtracting $8$ parameters. Furthermore, by making a change of basis in the space of solutions of the ODE, all monodromies  are conjugated by a constant matrix in $\textrm{SL}\left(3,\mathbb{R}\right)$ resp. $\textrm{SU}(1,2)$. Since  this conjugation generically acts effectively (i.e. it sweeps out an 8-dimensional subspace), we should subtract another 8 parameters, leading to the desired dimension  $16K - 4K - 8 - 8 = 12 K -16$.

Now we compute the dimension of the space of monodromy matrices after imposing the
$\textrm{SL}\left(3,\mathbb{R}\right)$ resp. $\textrm{SU}(1,2)$ conditions. The dimension of these real groups is 8, leading to $8K$ parameters a priori.  Again the invariants $\tr M_i$ and $\tr M_i^2$ are fixed in terms of the $\epsilon^{(s)}_i$, but now these invariants are automatically real, therefore subtracting $2K$ parameters. The reality constraint on the monodromy when encircling all points in the unit disk is now automatically satisfied, and the overall conjugation by a constant matrix in $\textrm{SL}\left(3,\mathbb{R}\right)$ resp. $\textrm{SU}(1,2)$ subtract $8$ parameters. This leads to a dimension of $8K-2K-8 = 6K -8$.

Computing the difference of these two dimensions leads to the number of  $6K-8$  constraints we need to impose to reduce the monodromy to  $\textrm{SL}\left(3,\mathbb{R}\right)$ resp. $\textrm{SU}(1,2)$, 
and matches precisely  the number of undetermined accessory parameters at our disposal.
We conclude that, generically, imposing the single-valuedness of the Toda fields will precisely fix all the accessory parameters.

\section{Monodromy problem for classical $\calw_3$  blocks}\label{secblocks}

In the previous section we have shown that constructing backreacted  bulk solutions containing certain spinning 
particles in the bulk higher spin theory reduces to solving a certain monodromy problem, where the accesssory parameters in an ODE are  fixed by requiring monodromies to be in 
SL$(3,\RR)$ resp. SU$(1,2)$.
Another context where a similar monodromy problem
appears, is in the construction of $\calw_3$ blocks at large central charge.
In this case, the monodromy properties of  the ODE are determined by the quantum numbers of the exchanged $\calw_3$ families. In this section we will show that the bulk monodromy problem is in fact identical to the one which determines a
$\calw_3$ vacuum block in a specific channel.
 We start with a brief discussion of the properties of $\mathcal{W}_N$ blocks.

\paragraph{$\mathcal{W}_3$ blocks}
As for the Virasoro algebra, the  $\mathcal{W}_N$ blocks are defined to be the building blocks of correlators which capture the purely  kinematical information which is
fixed by the $\mathcal{W}_N$ Ward identities.
The $\mathcal{W}_N$-blocks are much less studied objects than their Virasoro counterparts, see however \cite{Bowcock:1993wq}\cite{Wyllard:2009hg}\cite{Fateev:2011qa}\cite{Coman:2017qgv}. 

One peculiarity  of  $\calw_N$ blocks as opposed to Virasoro blocks goes back to the  well-known  fact that the  $\calw_N$ Ward identities don't allow  one to express arbitrary correlation functions in terms of correlators involving only $\calw_N$ primaries \cite{Bowcock:1993wq}.
This leads to the property that generic
 $\calw_N$ blocks depend on an infinite number of extra parameters besides the familiar dependence on  the chosen channel,  the cross-ratios and the quantum number of the external and exchanged primaries. This arbitrariness is however absent  for the vacuum blocks which will turn
 out to be  the ones relevant for our purposes.

Let us illustrate these features for the case of the four-point block (which will be relevant for constructing a two-centered bulk solution), and comment on the the generalization to higher point blocks at the end. 
We start by considering  a correlation function of four $\mathcal{W}_3$ primary operators\footnote{ In order to simplify equations, we  are displaying only the holomorphic coordinate dependence in our formulas.}$$\mathcal{A}_4 = \langle \calo_{\h_1} (z_1)  \calo_{\h_2} (z_2) \calo_{\h_3} (z_3)  \calo_{\h_4} (z_4) \rangle,$$ 
Here, $\mathcal{O}_\h$  is a $\mathcal{W}_3$-primary operator 
characterized by it's weights $\h = (h, w)$ with respect to $L_0$ and $W_0$.
The four-point function can be  decomposed into a sum over exchanged $\calw_3$-families
\begin{equation}
\mathcal{A}_4 = \sum_{\eta \in \{ \text{all primaries} \} } \langle \calo_{\h_1} (z_1)  \calo_{\h_2} (z_2) \Pi_{\eta}  \calo_{\h_3} (z_3)  \calo_{\h_4} (z_4) \rangle \,,
\end{equation}
where $\Pi_{\eta}$ is a projector onto a particular $\mathcal{W}_3$ descendant family of a primary state $|\eta \rangle = \mathcal{O}_{\eta}| 0 \rangle$,
\begin{equation}
\Pi_{\eta} = \sum_{I_{1,2},J_{1,2}}L_{-I_1}W_{-J_1}|\eta \rangle G^{J_1 I_1, I_2 J_2} \langle \eta| W_{J_2} L_{I_2}\,.
\end{equation}
Here 
capital $I$ is a multi-index so that $L_{I}$ is an abbreviation for $L_{i_1} L_{i_2} \ldots L_{i_k}$. The $G^{J_1 I_1, I_2 J_2}$ are elements of the  inverse of the inner product matrix $G_{J_1 I_1, I_2 J_2}=\langle \eta| W_{J_1} L_{I_1} L_{-I_2}W_{-J_2}|\eta \rangle$.
Mimicking the definition of Virasoro conformal blocks, we consider the ratio
\begin{equation} \label{conbl}
\calb_\h = { \langle \calo_{\h_1} (z_1)  \calo_{\h_2} (z_2) \Pi_\h  \calo_{\h_3} (z_3)  \calo_{\h_4} (z_4) \rangle \over 
\langle	 \calo_{\h_1} (z_1)  \calo_{\h_2} (z_2) | \h \rangle    \langle \h|  \calo_{\h_3} (z_3)  \calo_{\h_4} (z_4) \rangle}\,.
\end{equation}
However, unlike in the Virasoro case, the  quantity $\calb_\h$  is not completely fixed by $\calw_3$ kinematics and still depends on dynamical information. As shown in \cite{Bowcock:1993wq},\cite{Wyllard:2009hg}, $\calb_\h$ generically still depends on the ratios of three-point functions
\be 
C_n = {\langle \h_1 |  \calo_{\h_2} (1)   (W_{-1})^n | \h \rangle \over
\langle \h_1 |  \calo_{\h_2} (1)  |  \h \rangle}
, \qquad \bar C_n = {\langle \h |  (W_{1})^n  \calo_{\h_3} (1)  |  \h_4 \rangle  \over \langle \h |   \calo_{\h_3} (1)  |  \h_4 \rangle}.
\label{extrapars}
\ee 
One therefore defines the $\calw_3$ block to be an object depending on an extra set of 
free parameters $C_n, \bar C_n$, which
 in a specific $\calw_3$ CFT  should be specialized to take the appropriate values (\ref{extrapars}). It will be important to note that this infinite arbitrariness is absent\footnote{Another case where the
 arbitrariness is absent is when the external operators are semi-degenerate  \cite{Wyllard:2009hg}.} for the vacuum block, since
 the $C_n, \bar C_n$ vanish when 
 $|\h \rangle= |0 \rangle$ is the vacuum.

Summarizing, we have represented the correlation function $\mathcal{A}_4$ as
\begin{equation}
    \cala_4 = \sum_{\h}	 \langle	 \calo_{\h_1} (z_1)  \calo_{\h_2} (z_2) | \h \rangle    \langle \h|  \calo_{\h_3} (z_3)  \calo_{\h_4} (z_4) \rangle \calb_\h\, .
\end{equation}
Note that in writing this expansion, we have also chosen a `channel', i.e. we have chosen to fuse $\calo_{\h_1}$ with   $ \calo_{\h_2}$ and  $\calo_{\h_3}$ with   $ \calo_{\h_4}$.
Similarly, one can perform a decomposition of an $n$-point correlation function by inserting $n-3$ projectors for the exchanged primaries and dividing by $n-2$ three point functions, analogous to \eqref{conbl}.

\paragraph{Inserting a degenerate primary} 
As is the case for Virasoro blocks  (see e.g. \cite{Hartman:2013mia}) the
$\calw_N$ blocks in a certain large-$c$ limit can be determined by solving
a monodromy problem \cite{Coman:2017qgv}  \cite{deBoer:2014sna}. To derive
this, one considers the amplitude with an insertion of an additional degenerate
primary and makes use of the constraints imposed by   the decoupling of its null descendants.

To this end we consider the
 auxiliary object $\mathcal{B}_{\eta}[\Psi]$, which differs from the block 
 $\calb_\h$ defined in \eqref{conbl}  by an insertion of an extra degenerate primary $\Psi$ in the numerator:
\begin{equation}
\mathcal{B}_{\eta}[\Psi] = { \langle \Psi(z) \calo_{\h_1} (z_1)  \calo_{\h_2} (z_2) \Pi_\h  \calo_{\h_3} (z_3)  \calo_{\h_4} (z_4) \rangle \over
\langle	 \calo_{\h_1} (z_1)  \calo_{\h_2} (z_2) | \h \rangle    \langle \h|  \calo_{\h_3} (z_3)  \calo_{\h_4} (z_4) \rangle}.\label{auxil}
\end{equation}
Here, $\Psi$ is a specific degenerate operator  associated to a state which satisfies shortening conditions of the form:
\begin{equation} \label{degg}
\begin{split}
\left[ L_{-1} + \kappa_{1,1} W_{-1} \right] |\Psi \rangle &=0 \,,
\\
\left[ L^2_{-1} + \kappa_{2,1} L_{-2} + \kappa_{2,2} W_{-2}  \right] |\Psi \rangle &= 0 \,,
\\
\left[ L^3_{-1} + \kappa_{3,1} L_{-2} L_{-1}  + \kappa_{3,2} L_{-3}  + \kappa_{3,3} W_{-3}  \right] |\Psi \rangle &= 0\,.
\end{split}
\end{equation}
The explicit coefficients\footnote{In what follows we will only need the large $c$ behaviour 
$\kappa_{3,1} \approx 2 \kappa_{3,2} \approx
\kappa_{3,3} \approx  {24 \over c}$}
can be found in \cite{Coman:2017qgv}. Upon inserting the last  equation into the correlation function (\ref{auxil}), we obtain a shortening relation  of the form
\begin{equation} \label{dege}
    \left[ \partial^3   + \kappa_{3,1} \hat \calt (z) \partial  +\kappa_{3,2}  \hat \calt'(z) + \kappa_{3,3}  \hat \calw (z) \right] \mathcal{B}_{\eta}[\Psi] = 0 \,.
\end{equation}
The operators $\hat \calt, \hat \calw$ which act on the  $\mathcal{B}_{\eta}[\Psi]$ are defined as
\begin{eqnarray}
       \hat \calt &=& \sum_i \frac{h_i}{(z-z_i)^2} + \frac{1}{(z-z_i)} \frac{\partial \  }{\partial z_i} \,,
        \\
       \hat  \calw &=& \sum_i \frac{w_i}{(z-z_i)^3} + \frac{W^{(i)}_{-1}}{(z-z_i)^2} + \frac{W^{(i)}_{-2}}{(z-z_i)} \,,
\end{eqnarray}
where  $h_i$ and $w_i$ are  the conformal dimension and spin-3 charge of the $i^{\rm th}$ operator. By $W^{(i)}_{-k}$ we denote the negative $k^{\rm th}$ mode acting on $i^{\rm th}$ operator inside the $\mathcal{B}_{\eta}[\Psi]$, for example   $W^{(1)}_{-k}\mathcal{B}_{\eta}[\Psi]$ is shorthand for the ratio
\begin{equation}
    W^{(1)}_{-k}\mathcal{B}_{\eta}[\Psi]
    =
    \frac{ \langle\Psi(z) \left( W_{-k} \calo_{\h_1} \right) (z_1)  \calo_{\h_2} (z_2) \Pi_\h  \calo_{\h_3} (z_3)  \calo_{\h_4} (z_4) \rangle }{
\langle	 \calo_{\h_1} (z_1)  \calo_{\h_2} (z_2) | \h \rangle    \langle \h|  \calo_{\h_3} (z_3)  \calo_{\h_4} (z_4) \rangle} \,.
\end{equation}
As it stands, equation (\ref{dege}) is a  differential equation which couples different amplitudes. We will now argue that, in a suitable large-$c$ limit,
it reduces to an ODE for a single function.

\paragraph{Classical, large $c$, limit} 
We now consider the limit of large central charge $c$, in which the  primary
operators $\calo_{\h_i}$ are assumed to be \qq{heavy} in the sense that $ \eta_i/c$
remains finite in the limit.  
It has been argued, see e.g. \cite{Harlow:2011ny}, although not proven, that in this limit the general conformal block exponentiates 
\begin{equation}
     \lim_{c \rightarrow \infty } \mathcal{B}_{\eta} = \ex^{ -\frac{c}{6} b_{\nu} } \,.
     \label{bkgd}
\end{equation}
Here, the parameters  $\nu_i = (\e_i, \d_i) $ on the right hand side contain the rescaled \qq{classical} weights
\begin{equation}
    \epsilon_i = \frac{24}{c} h_i,
    \qquad
    \delta_i = \frac{24}{c} w_i\, .\label{rescaled}
\end{equation}
The behaviour of (\ref{bkgd}) is reminiscent of a saddle-point approximation, where 
$b_\n$ is the action of the saddle-point.

Now let's consider the quantity $\mathcal{B}_{\eta} [\Psi]$ in (\ref{auxil}) with 
the insertion of the degenerate primary $\Psi$. Since $\Psi$ is \qq{light}, in the sense that its charges are of order 1 at large $c$, it is natural to assume that 
its presence does not change the saddle point and that 
$\mathcal{B}_{\eta} [\Psi]$ factorizes as
\begin{equation}
     \lim_{c \rightarrow \infty } \mathcal{B}_{\eta} [\Psi] = \psi_{\nu} \ex^{ -\frac{c}{6} b_{\nu}} \,. \label{wavef0}
\end{equation}
We expect a similar factorization to occur in the case of $W^{(i)}_{-k}\mathcal{B}_{\eta} [\Psi]$ since the action of $W^{(i)}_{-k}$ does not change the leading behaviour of $\mathcal{W}_3$ weights in the large $c$ limit,
so that
\begin{equation}
     \lim_{c \rightarrow \infty }{ W^{(i)}_{-k}\mathcal{B}_{\eta} [\Psi]
     \over  W^{(i)}_{-k}\mathcal{B}_{\eta}} = \psi_{\nu}  \,. \label{wavef}
\end{equation}
An important assumption here is  that this factorization involves the same function 
$\psi_{\nu}$ as in (\ref{wavef}), see \cite{Harlow:2011ny} for a justification.

Under these assumptions,
the decoupling equation (\ref{dege}) reduces in the
large-$c$ limit to at holomorphic ODE for  the \qq{wave function} $\psi_\n$
\begin{equation}
   \left( \partial^3  + \calt \partial  + \frac{1}{2} \partial \calt  +\calw \right) \psi_{\nu} = 0\,. \label{ODECFT}
\end{equation}
Here $\calt ,\calw$ are the functions
\begin{eqnarray}
\calt (z) &=& \sum_{i=1} \left({\e_i \over (z-z_i)^2} + {c_i \over (z-z_i)}\right) \,,
\\
\calw (z) &=& \sum_{i=1} \left({\d_i \over (z-z_i)^3} + {d_i \over (z-z_i)^2 } +  {a_i \over (z-z_i) }\right) \,,
\end{eqnarray}
with coefficients $\epsilon_i,\delta_i$ defined in (\ref{rescaled}), and the accessory parameters $a_i,d_i,c_i$ given by
 the following limits
\begin{eqnarray} \label{coef}
c_i =  \lim_{c \to \infty} \frac{24}{c} \frac{L^{(i)}_{-1} \mathcal{B}_{\eta}}{ \mathcal{B}_{\eta} },
\qquad
d_i =  \lim_{c \to \infty} \frac{24}{c} \frac{W^{(i)}_{-1} \mathcal{B}_{\eta}}{ \mathcal{B}_{\eta} },\quad \text{and}
\quad
a_i =  \lim_{c \to \infty} \frac{24}{c} \frac{W^{(i)}_{-2} \mathcal{B}_{\eta}}{ \mathcal{B}_{\eta} }\, .
\end{eqnarray} 
The exponentiation of the classical block (\ref{bkgd}) implies that the $c_i$ are finite
and given by
\be 
c_i = - 4 \pa_{z_i} b_\n .\label{cfromblock}
\ee
It is natural to expect that the $a_i$ and $d_i$  similarly remain  finite in the limit.

To summarize, we have found associated to a classical $\calw_3$ block an ODE (\ref{ODECFT}) which is of the same form as the auxiliary ODE (\ref{psieq}) determining the solutions of the $\cala_2$ Toda
system. 
Similarly to what happens for classical Virasoro blocks \cite{Hartman:2013mia}, the choice of the exchanged $\calw_3$ primary  $\h$ determines  the monodromy properties of the ODE (\ref{ODECFT}). To see this, note
that the quantity (\ref{auxil}) contains the four-point function
\be 
\langle   \calo_{\h_1} (z_1)  \calo_{\h_2} (z_2) \Psi(z) |\h \rangle  =  \langle   \calo_{\h_1} (z_1)  \calo_{\h_2} (z_2) \Psi(z) \calo_\h (0) \rangle \,.
\ee
Using the OPE between the light operator $\Psi(z)$ and the heavy operator
$\calo_\h (0) $ one can show that the trace invariants $\tr M, \ tr M^2$  of the monodromy matrix  as $\Psi(z)$ encircles the origin
 are fixed in terms of $\h$, and this is of course also the same monodromy when $z$ encircles both $z_1$ and $z_2$.
This discussion generalizes straightforwardly  to  classical $n$-point blocks:
they are similarly determined by an ODE whose monodromy properties are determined by
the choice of exchanged primaries in the chosen channel.

\paragraph{Relating the monodromy problems}

Now we are ready to relate the  monodromy problem for the classical $\calw_3$  four-point block to the  monodromy  problem which determines a 2-centered solution  in  spin-3 gravity.
In studying the latter problem we  
 encountered an ODE on the complex plane of the form   (\ref{ODECFT})
with identical singularities in two pairs of image points. Therefore we consider 
a CFT four-point function with a  primary $\calo_{\h_1}$  inserted in the points $z_1$ and $1/ \bar z_1$, and a second primary $\calo_{\h_2}$  inserted in $z_2$ and $1/ \bar z_2$.
Next we should choose a channel in which to perform the conformal block expansion. 
It turns out that the relation between the two monodromy problems is the simplest if we choose the \qq{mirror channel} in which we fuse the operators located in image points (i.e. $z_1$ and $1/ \bar z_1$ resp. $z_2$ and $1/ \bar z_2$)  together in pairs. 
The reason is that we derived  in paragraph (\ref{imagemon}) above that, in the bulk problem, the monodromy when encircling a pair of image points is the identity. 
 From our discussion in the previous paragraph, this means that 
  the exchanged primary in the corresponding  
  $\calw_3$ block in the mirror channel  is the identity operator. In other words, our bulk monodromy problem determines a  $\calw_3$ vacuum block.

\begin{figure} 
    \centering
    \includegraphics{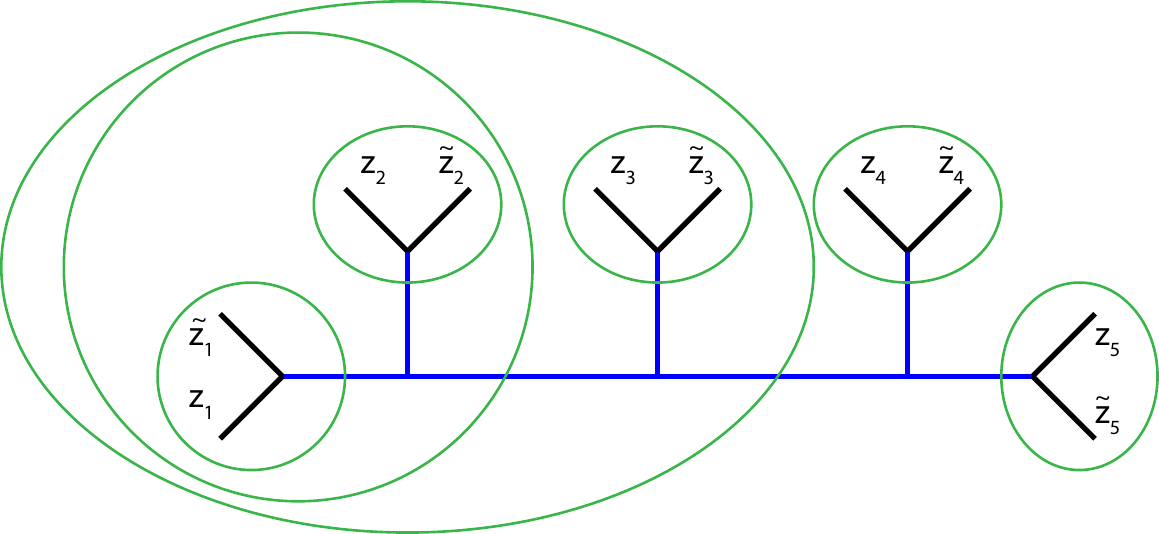}
    \caption{
    The diagram represents a $10$-point conformal block on a sphere . We use black legs for the
    external primaries inserted at points $z_i$ and their images $\tilde{z}_i = \frac{1}{\bar{z}_i}$. The blue lines correspond to the exchange of the identity family. This specific channel, where the mirror pairs are fused first, is the one related to the monodromy problem on a disk. The green circles show contours along which the monodromy is trivial. The picture is taken from \cite{Hulik:2016ifr}.
    }
    \label{fig1}
\end{figure}

In summary, we have argued that the monodromy problem determining  a 2-centered solution  in  spin-3 gravity is equivalent to that which determines the classical vacuum 4-point block
\begin{equation}
    b_0 = - \lim_{c \to \infty}  {6 \over c} \ln
    \frac{ 
    \langle  
    \calo_{\h_1} (z_1)  \calo_{\h_1} (\frac{1}{\bar{z}_1}) 
    \Pi_0
    \calo_{\h_2} (z_2)  \calo_{\h_2} (\frac{1}{\bar{z}_2}) 
    \rangle
    }{
\langle	 \calo_{\h_1} (z_1)  \calo_{\h_1} (\frac{1}{\bar{z}_1})  | 0 \rangle    
\langle 0 | \calo_{\h_2} (z_2)  \calo_{\h_2} (\frac{1}{\bar{z}_2}) \rangle} \,.
\end{equation}
 The argument  can be generalized in a straightforward manner to show that the monodromy problem for a $K$-centered solution determines a classical vacuum $2K$-point block in a \qq{mirror} channel where the operators in image points are fused together in pairs (see fig.\ref{fig1}).
We should stress that, due to the fact that  $\calw_3$ vacuum blocks are unique and  don't depend on extra parameters like (\ref{extrapars}),
the derived correspondence is really one-to-one:
 solving
the monodromy problem for the $K$-centered solution determines the 2K-point
vacuum block through (\ref{cfromblock}) and,
conversely, from the knowledge of the vacuum block we can in principle derive the accessory parameters in the ODE  and construct the bulk multi-centered solution.
\footnote{Unlike for $c_i$, for the parameters $a_i$ and $d_i$ no closed form expression like \eqref{cfromblock} exists. So it seems that in order to extract them from \eqref{coef} one needs not only $\mathcal{B}_0$ but also $W^{(i)}_k \mathcal{B}_0$. These however, for the case of the vacuum conformal block, can in principle be derived from $\mathcal{B}_0$ using $\mathcal{W}_3$ ward identities and the properties of $\mathcal{W}_3$ vacuum.}

\section{Generalization to Arbitrary Spin}\label{secspinN}

Here we present generalization of the above results to higher spin theories with spins $2, \ldots , N$. Our results in this section are either based on generic arguments or are extrapolated from explicit calculations we carried out for $N=3$  and $N=4$. Many of the details of Toda theory related to this section are discussed in appendix \ref{TodaTheory}. We focus on gauge connections that belong either to the maximal, $\textrm{sl}(N,\mathbb{R})$, or to the next to maximal, $\textrm{su}(\lfloor \frac{N}{2} \rfloor , \lceil\frac{N}{2}\rceil)$, non-compact real form of $\textrm{sl}(N,\mathbb{C})$.

\subsection{The Toda gauge connection}

First we observe that for $N=3$, the left connection \eqref{ATsl3} in Toda gauge can be rewritten for both values of $\sigma$ as 
\begin{equation}\label{ATsl3both}
    \begin{split}
        A_T = & \frac{1}{2}\left(\partial\f_1\dif z - \bar{\partial}\f_1\dif\bar{z}\right)H_1 + \ex^{-\f_1+\frac{1}{2}\f_2}\left(\im E_1^-\dif z - \im E_1^+\dif\bar{z}\right) \\
        & + \frac{1}{2}\left(\partial\f_2\dif z - \bar{\partial}\f_2\dif\bar{z}\right)H_2 + \ex^{-\f_2+\frac{1}{2}\f_1}\left(\im E_2^-\dif z - \im E_2^+\dif\bar{z}\right) \, ,
    \end{split}
\end{equation}
where the matrices $H_i$ and $E^{\pm}_i$ are defined in appendix \ref{connectionbasis}. From \eqref{ATsl3both} we can generalize to arbitrary spin by writing the connection 
\begin{equation}\label{slnconnection1}
    A_T =   \frac{1}{2}\left(\partial\Phi \dif z - \bar{\partial}\Phi\dif\bar{z}\right) + \im E_i^-e^{-\frac{1}{2}\alpha_i(\Phi)}\dif z -\im  E_i^+e^{-\frac{1}{2}\alpha_i(\Phi)}\dif\bar{z}\, ,
\end{equation}
where $\Phi = H_i\phi^i$. 
Then \eqref{slnconnection1} becomes
\begin{equation}\label{slnconnection2}
    A_T =  \frac{1}{2}\left(\partial\phi^i \dif z - \bar{\partial}\phi^i\dif\bar{z} \right)H_i + \im E_i^-e^{-\frac{1}{2}C_{ij}\phi^j}\dif z -\im  E_i^+e^{-\frac{1}{2}C_{ij}\phi^j}\dif \bar{z}\, ,
\end{equation}
where $C_{ij}= \alpha_i (H_j)$ is the Cartan matrix and is obtained from
\begin{equation}\label{simpleroots}
    \left[H_i,E_j^{\pm}\right] = \pm\alpha_j\left(H_i\right)E_j^{\pm}\, .
\end{equation}
In matrix form we have
\begin{equation}\label{Cmatrix}
    C =  \begin{pmatrix}
2 & -1 & 0 & \ldots
\\
-1 & 2 & -1 & \ldots
\\
0 & -1 & 2 & \ldots
\\
\vdots&\vdots&\vdots& \ddots
\end{pmatrix}\, .
\end{equation}
The flatness of the $A_T$ is equivalent to the Toda system being satisfied.
\begin{equation}\label{ANTodaEqnsMain}
    \begin{matrix}
    \partial\bar{\partial}\phi_1+ e^{-2\phi_1+\phi_2}=0 \, ,\\
    \partial\bar{\partial}\phi_2+ e^{-2\phi_2+\phi_1+\phi_3}=0 \, ,\\
    \vdots\\
    \partial\bar{\partial}\phi_j+ e^{-2\phi_j+\phi_{j-1}+\phi_{j+1}}=0\, ,\\
    \vdots\\
    \partial\bar{\partial}\phi_{N-1}+ e^{-2\phi_{N-1}+\phi_{N-2}}=0\, .
    \end{matrix}
\end{equation}
The argument to be made is that \eqref{slnconnection2} can be brought into a form similar to \eqref{ATsl3} with the presence of a parameter $\sigma$ such that
\begin{align}
& \tilde{A}\in \textrm{sl}(N,\RR),\  &\s=& 1: &  \f_i &= \bar \f_{N-i}\, , \nonu
&\tilde{A}\in \textrm{su}\left(\lfloor \frac{N}{2} \rfloor,\lceil\frac{N}{2}\rceil\right),\ &\s=& -1: &  \f_i &\in \RR.\label{realcondsN}
\end{align}
Similarly we choose the right connection $\bar{A}_T$ to be proportional to the principally embedded Cartan element of $\textrm{sl}(2,\RR)$ inside $\textrm{sl}\left(N,\RR\right)$ respectively $\textrm{su}\left(\lfloor \frac{N}{2} \rfloor , \lceil\frac{N}{2}\rceil\right)$ 
\begin{equation}\label{generalAbar}
\bar{A}_T = \im\frac{1}{2}\sum_i i (N-i)H_i \dif\tilde{t} = \im\bar{L}_0 \dif\tilde{t}\, ,
\end{equation}
which is trivially flat.

As discussed in appendix \ref{WilsonParticles} in the presence of backreacting point particles the equations of motion become
\begin{equation}\label{spinNeoms}
    \begin{matrix}
    \partial\bar{\partial}\phi_1+ e^{-2\phi_1+\phi_2} =\frac{\pi}{k}\sum_i \alpha^{(i)}_1\delta^2\left(z-z_i,\bar{z}-\bar{z}_i\right) \, ,\\
    \partial\bar{\partial}\phi_2+ e^{-2\phi_2+\phi_1+\phi_3}  =\frac{\pi}{k}\sum_i \alpha^{(i)}_2\delta^2\left(z-z_i,\bar{z}-\bar{z}_i\right) \, ,\\
     \vdots\\
    \partial\bar{\partial}\phi_i+ e^{-2\phi_j+\phi_{j-1}+\phi_{j+1}}=\frac{\pi}{k}\sum_i \alpha^{(i)}_j\delta^2\left(z-z_i,\bar{z}-\bar{z}_i\right)\, ,\\
    \vdots\\
    \partial\bar{\partial}\phi_{N-1}+ e^{-2\phi_{N-1}+\phi_{N-2}}=\frac{\pi}{k}\sum_i \alpha^{(i)}_{N-1}\delta^2\left(z-z_i,\bar{z}-\bar{z}_i\right)\, .
    \end{matrix}
\end{equation}
where for the $i$-th particle we assumed the decomposition of the momentum
\begin{equation}
P_{L,i} = \sum_j \alpha_j^{(i)}H_j\, .
\end{equation}
The constants $\alpha_j^{(i)}$ can be expressed in terms of the quantum numbers of the particles from the constraints coming from the Lagrange multipliers, as explained in appendix \ref{WilsonLines}.

As explained in appendix \ref{TodaTheory} there are two systems of associated ODEs \eqref{TodaANU1} and \eqref{TodaANU2} expressed in terms of holomorphic and antiholomorphic currents $U^{(s)}_{1,2}$, $V^{(s)}_{1,2}$ from which we can construct (quasi-)primary currents $\mathcal{W}^{(s)}$, where $s=2,\ldots , N$.

\subsection{Properties of the currents}\label{currentsproperties}

\paragraph{Pole structure:}
If near the sources $z_i$ we neglect the potential term compared to
the kinetic term, then the Toda fields $\f_i$ behave near $z = z_i$ as
\begin{equation}
    \f_j\sim \frac{\alpha_{j}^{(i)}}{k}\ln |z-z_i|\, ,
\end{equation}
Thus the solution of the Toda equations on the disk specifies meromorphic (quasi-)primary currents $\mathcal{W}^{(s)}$ with $s=2,\ldots,N$. The sources $\alpha_j^{(i)}$ fix the most singular terms in $\mathcal{W}^{(s)}$ which take the form
\begin{equation}
\mathcal{W}^{(s)}(z)=\sum_{i=1}^K\frac{\epsilon^{(s)}_i}{\left(z-z_i\right)^s} + \ldots \, ,
\end{equation}
where the ellipses denote lower order poles and a regular part. The constants $\epsilon^{(s)}_i$ are specified in terms of $\alpha^{(i)}_j$ (with $j=i,\ldots,N-1$) by examining the form of the Toda fields $\phi_j$ near a specific source.

\paragraph{Doubling trick:} As shown for the $N=3$ case
\eqref{realcurrents}, the currents $\mathcal{W}^{(s)}(z)$ must satisfy the following reality condition on the boundary circle
\begin{align}
&\textrm{sl}(N,\RR),\  \s= 1:&  \left.\left( \im^s z^s  \calw^{(s)} \right)\right \rvert_{|z|=1} \in \RR, \nonu
&\textrm{su}\left(\lfloor \frac{N}{2} \rfloor , \lceil\frac{N}{2}\rceil\right),\ \s= -1:& \left.\left( z^s  \calw^{(s)} \right)\right \rvert_{|z|=1} \in \RR.\label{realcurrentsN}
\end{align}
The reality condition is imposed by using a ‘doubling trick’ and extending $\mathcal{W}^{(s)}(z)$ to meromorphic functions on the complex plane in a suitable manner. Using the Schwarz reflection principle we find that the appropriate reflection properties are
\begin{equation}\label{reflectionproperty}
\mathcal{W}^{(s)}(z) = \frac{\left(-\sigma\right)^s}{z^{2s}}\bar{\mathcal{W}}^{(s)}\left(\frac{1}{z}\right)\, .
\end{equation}

\paragraph{Constraints on the accessory parameters:} As a result of (\ref{reflectionproperty}) $\mathcal{W}^{(s)}$ now has poles both at the $z_i$ and their image points $\bar{z}_i^{-1}$. Thus $\mathcal{W}^{(s)}$ is of the form
\begin{equation}\label{WsPoles}
\mathcal{W}^{(s)}(z)  =\sum_{i=1}^K\left(\frac{\epsilon^{(s)}_i}{\left(z-z_i\right)^s} + \frac{\tilde{\epsilon}^{(s)}_i}{\left(z-1/\bar{z}_i\right)^s} +\sum_{l=s-1}^{1}\left(\frac{c_i^{(s,l)}}{\left(z-z_i\right)^l} +\frac{\tilde{c}_i^{(s,l)}}{\left(z-1/\bar{z}_i\right)^l} \right)\right) \, ,
\end{equation}
where in the above equation we assumed for simplicity there is no source located at the origin and as a result there is also no image source at infinity. As in the $\textrm{sl}(3)$ case the parameters $c_i^{(s,l)}, \tilde{c}_i^{(s,l)}$ are the accessory parameters and are not determined by the $\alpha^{(i)}_j$. Instead they are determined by solving a monodromy problem, which arises from demanding that the Toda fields are single-valued. Once again not all of the parameters of \eqref{WsPoles} are independent since we have to impose the reflection condition \eqref{reflectionproperty}. For each of the currents $\mathcal{W}^{(s)}$ we have in total $4sK$ real parameters from which $4(s-1)K$ are accessory parameters, where $K$ denotes the number of insertions within the unit disk. Substituting (\ref{WsPoles}) in (\ref{reflectionproperty}) and requiring the equality of poles at each order near the $z_i$ we get $2sK$ real conditions. $2K$ out of them are
\begin{equation}\label{stildeeconstraint}
e^{(s)}_i = (\sigma)^s\bar{\tilde{e}}^{(s)}_i \, .
\end{equation}
Thus the parameters $\tilde{e}^{(s)}_i$ are also determined in terms of $\alpha^{(i)}_j$. So we are left with $4(s-1)K$ accessory parameters and $2(s-1)K$ real conditions on them which gives $2(s-1)K$ parameters left. Substituting these real conditions in (\ref{reflectionproperty}) and requiring regularity at the origin we get another $2s-1$ additional constraints.
Thus from each $\mathcal{W}^{(s)}(z)$ we get
\begin{equation}\label{sfreeparameters}
2(s-1)K-(2s-1) \, ,
\end{equation}
parameters. Summing up $s$ from $2$ to $N$ we get that from all the currents $\mathcal{W}^{(s)}$ we have in total
\begin{equation}\label{totalfreeparameters}
KN(N-1)-(N^2-1)\, ,
\end{equation}
parameters left that should be fixed by solving the monodromy problem.

\subsection{Doubling trick for $\psi_i$}
To prove the doubling trick for $\psi_i$ one needs the associated ODE expressed in terms of the (quasi-)primary currents $\mathcal{W}^{(s)}$. That however is not known for general $N$ and has to be worked out case by case. Thus here we will only sketch the general steps of the proof for arbitrary $N$, which justify our choice when it comes to the doubling trick of $\psi_i$.

Under a conformal transformation $z\rightarrow \tilde{z}=f(z)$ the fields in the associated ODE for $\psi_i(z)$ transform as
\begin{equation}
    \tilde{\mathcal{W}}^{(s)}\left(\tilde{z}\right) = \left(f'\right)^{-s}\left(\mathcal{W}^{(s)}\left(z\right) - \delta_{s,2}\beta_NS\left(f,z\right)\right) \, , \quad \tilde{\psi}_i\left(\tilde{z}\right) = \left(f'\right)^{\frac{N-1}{2}}\psi_i\left(z\right)\, .
\end{equation}
The transformation of $\psi_i$ above was derived from the transformation of $\phi_1$ \eqref{phiconftrans} and the fact that $\psi_i$ is the holomorphic part of $\ex^{\f_1}$. These transformations should leave the associated ODE conformally invariant. Upon performing a conformal transformation with $f(z)=1/z$ and using the reflection property \eqref{reflectionproperty} we expect to get an ODE with the presence of the parameter $\sigma$ from which one can deduce the following.

For the $\textrm{su}\left(\lfloor \frac{N}{2} \rfloor , \lceil\frac{N}{2}\rceil\right)$ case we have $\sigma=-1$ and thus
\begin{equation}
    \Psi(z) = S z^{N-1}\bar{X}\left(\frac{1}{z}\right)\, ,
\end{equation}
with $X(z)$ given by \eqref{chiMinorpsi}. Then
\begin{equation}\label{sunreflectionpsi}
    \psi_a(z) = S \epsilon_{ab_1\ldots b_{N-1}}\Lambda^{b_1\tilde{b}_1}\ldots\Lambda^{b_{N-1}\tilde{b}_{N-1}}\bar{\psi}_{\tilde{b}_1}\left(1/z\right)\partial_{\frac{1}{z}}\bar{\psi}_{\tilde{b}_2}\left(1/z\right)\ldots \partial_{\frac{1}{z}}^{N-2}\bar{\psi}_{\tilde{b}_{N-1}}\left(1/z\right)\, .
\end{equation}
For the $\textrm{sl}\left(N,\RR\right)$ case we have $\sigma=1$ and thus
\begin{equation}\label{slnreflectionpsi}
    \Psi(z) = \tilde{S} z^{N-1}\bar{\Psi}\left(\frac{1}{z}\right)\, .
\end{equation}
The matrices $S$ and $\tilde{S}$ are proportional to the identity and are respectively given by \eqref{SNmatrix} and \eqref{tildeSNmatrix}.

\subsection{Monodromies}
The monodromies  work out similarly to the $\textrm{SL}\left(3,\mathbb{C}\right)$ case. Thus for this section we mostly state the results as the proof is a direct generalization of what was described in section \ref{subsectionsl3monodromy}.
In general after encircling a singular point $z_i$ the solution transforms as
\begin{equation}
    \Psi \rightarrow M_i \Psi\, ,
\end{equation}
where $M_i\in \textrm{SL}(N,\mathbb{C})$ is a monodromy matrix. However requiring the Toda fields to be single valued constrains the monodromy matrix. Specifically for real Toda fields, since we have
\begin{equation}
    \ex^{\phi_1} = \Psi^{\dagger}\Lambda\Psi\, ,
\end{equation}
single valuedness requires that
\begin{equation}
    M_i^{\dagger}\Lambda M_i = \Lambda\, ,
\end{equation}
with $\Lambda = \textrm{diag}\left(1,-1,1,-1,\ldots\right)$. This means that we must have $M_i\in \textrm{SU}(\lfloor \frac{N}{2} \rfloor , \lceil\frac{N}{2}\rceil)$.\\
For complex conjugate Toda fields we have
\begin{equation} \label{sln}
    \ex^{\f_1}\sim \epsilon^{ab_1\ldots b_{N-1}}\psi_a\bar{\psi}_{b_1}\bar{\partial}\bar{\psi}_{b_2}\ldots \bar{\partial}^{N-2}\bar{\psi}_{b_{N-1}}\, .
\end{equation}
Requiring the field to be single valued and by using the definition of the adjugate matrix we get
\begin{equation}
    M_i=\bar{M}_i\, ,
\end{equation}
which implies that $M\in\textrm{SL}\left(N,\RR\right)$.

Using the reflection property of $\Psi$, \eqref{slnreflectionpsi} resp. \eqref{sunreflectionpsi} one can show that, provided the monodromy matrices belong to $\textrm{SL}\left(N,\RR\right)$ resp. $\textrm{SU}(\lfloor \frac{N}{2} \rfloor , \lceil\frac{N}{2}\rceil)$, the monodromy matrix of a contour encircling both a point and its image is the identity matrix. In a similar manner one can show that the monodromy matrix of a contour encircling all points within the unit disk is an element of $\textrm{SL}\left(N,\RR\right)$ resp. $\textrm{SU}(\lfloor \frac{N}{2} \rfloor , \lceil\frac{N}{2}\rceil)$.

\paragraph{Parameter counting and monodromy reduction}
Similarly to section \ref{parcunt} we will show that the condition the monodromy matrices take values in ${\rm SL}(N,\mathbb{R})$, respectively $\textrm{SU}( \lfloor \frac{N}{2} \rfloor , \lceil\frac{N}{2}\rceil)$ imposes in general precisely as many constraints as the number of available undetermined accessory parameters in the ODE. 
Once again we first compute the dimension of the space in which the monodromy matrices around the $K$ singular points in the unit disk take values and subtract from this the  dimension of the space in which they take values when the ${\rm SL}(N,\mathbb{R})$, resp. $\textrm{SU}( \lfloor \frac{N}{2} \rfloor , \lceil\frac{N}{2}\rceil)$ condition is imposed.

In general the monodromy matrix $M_i$ of a single point is an element of $\textrm{SL}\left(N,\mathbb{C}\right)$. Thus for the monodromy matrices of all $K$ points we have in total $2K(N^2-1)$ real parameters. However not all of these parameters are independent. The conjugacy class of each matrix $M_i$ is fixed due to the relation of the class functions $\textrm{tr}\left(M_i^k\right)$ (for $k=1,\ldots ,N-1$) to the higher spin charges of the particle. The latter subtracts $2K(N-1)$ parameters. Furthermore the monodromy matrix, when encircling all singular points in the unit disk, must take values in one of the real forms ${\rm SL}(N,\mathbb{R})$, resp. $\textrm{SU}( \lfloor \frac{N}{2} \rfloor , \lceil\frac{N}{2}\rceil)$, thus subtracting $N^2-1$ parameters. The last constraint comes from the conjugation of all monodromies by a constant matrix, in ${\rm SL}(N,\mathbb{R})$, resp. $\textrm{SU}( \lfloor \frac{N}{2} \rfloor , \lceil\frac{N}{2}\rceil)$, after making a change of basis in the space of solutions of the ODE. Thus we should subtract another $N^2-1$ parameters. This leads to $2KN(N-1)-2(N^2-1)$ independent parameters.

Next we compute the dimension of the space of monodromy matrices after imposing the  ${\rm SL}(N,\mathbb{R})$, resp. $\textrm{SU}( \lfloor \frac{N}{2} \rfloor , \lceil\frac{N}{2}\rceil)$ conditions. These real forms have dimentions $N^2-1$, leading to $K(N^2-1)$ parameters. Again the conjugacy class is fixed, but now the invariants $\textrm{tr}\left(M_i^k\right)$ are automatically real, therefore subtracting $K(N-1)$ parameters. The reality constraint on the monodromy when encircling all points in the unit disk is now automatically satisfied, and the overall conjugation by a constant matrix in ${\rm SL}(N,\mathbb{R})$, resp. $\textrm{SU}( \lfloor \frac{N}{2} \rfloor , \lceil\frac{N}{2}\rceil)$ subtracts $N^2-1$ parameters. This leads to a dimension of $KN(N-1)-(N^2-1)$.

Computing the difference of these two dimensions leads to the number of  $KN(N-1)-(N^2-1)$  constraints we need to impose to reduce the monodromy to  ${\rm SL}(N,\mathbb{R})$, resp. $\textrm{SU}( \lfloor \frac{N}{2} \rfloor , \lceil\frac{N}{2}\rceil)$, and matches precisely  the number of undetermined accessory parameters at our disposal. Thus, generically, imposing the single-valuedness of the Toda fields will precisely fix all the accessory parameters and guarantees that for generic values of the particle positions and quantum numbers, our equations have a unique solution.

\paragraph{Relation to classical $\calw_N$ blocks}
Our  arguments in section \ref{secblocks} for the $\calw_3$ case can be generalized in a straightforward manner to show that the monodromy problem determining a $K$-centered solution in 
the spin-$N$ theory is equivalent to the one determining a classical vacuum $2K$-point $\calw_N$ block in a \qq{mirror} channel where the operators in image points are fused together in pairs.

\section{Outlook}

To conclude, we list some open problems and possible generalizations:
\begin{itemize}
    \item In this work we argued for a connection between backreacted multi-centered solutions and $\calw_N$ blocks from studying the bulk equations of motion. It would also be of interest to evaluate the (regularized) bulk action on these solutions, which one would expect, based on results in the heavy-light approximation \cite{Hijano:2015rla}, to compute the $\calw_N$ vacuum block $\calb_0$ itself. Such an analysis would extend the holographic computation of Virasoro and $\calw_N$ blocks beyond the
    heavy-light approximation.
    \item We restricted our attention to the description of a rather special class of particles localized in the bulk, which source only the left-moving Chern-Simons connection. It would be of obvious interest to generalize this to the case where both  left- and right-moving connections are excited and make contact with other approaches to bulk localization \cite{Verlinde:2015qfa}, \cite{Nakayama:2015mva}.
    \item As we remarked before in section \ref{secgrav}, the configurations we studied in this work are particular to the Lorentzian bulk theory, because the geodesics on which our particles move do not analytically continue to geodesics in  Euclidean AdS. In contrast, geodesics in the Euclidean AdS begin and end on the boundary, and are described by localized  excitations on the boundary. It seems likely that this Euclidean setup has a simple description in terms of the free field variables on the boundary introduced in \cite{Campoleoni:2017xyl}.
    \item In our multi-centered solutions, we restricted the individual centers to be  particles rather than black holes. It would be desirable to generalize our approach  to include black hole centers.  It would also be worthwhile to construct solutions where the individual centers are particles but the monodromy when encircling all centers is that of a black hole.  These would be toy models for black hole microstates in which questions about information loss could be addressed \cite{Fitzpatrick:2016ive}.
    \item The real forms appearing in our discussion are $\mathrm{SL}(N,\mathbb{R})$ and $\textrm{SU}( \lfloor \frac{N}{2} \rfloor , \lceil\frac{N}{2}\rceil)$. However, one might wonder whether more general $\textrm{SU}(p,q)$'s are physically relevant. There is seemingly only one restriction given by the existence of $\textrm{SL}( 2, \mathbb{R})$ embeddeding in the $\textrm{SU}(p,q)$ which corresponds to the gravitational subsector. A full answer would requre a thorough study of reality conditions of Toda fields and appropriate $\textrm{SL}( 2, \mathbb{R})$ embeddings. Secondly, an investigation of embeddings other than the principal one might also be of interest.
\end{itemize}

\acknowledgments
We would like to thank Kara Farnsworth, Renann L. Jusinskas, Tomas Prochazka, Monica Guica, Elli Pomoni, Massimo Porrati, Jan de Boer and Erik Perlmutter for valuable discussions. The research of O.H., J.R. and O.V. was supported by the Grant Agency of the Czech Republic under the grant 17-22899S.  The research of J.R. and O.V. was supported by ESIF and MEYS (Project CoGraDS - CZ.02.1.01/0.0/0.0/15
003/0000437).  O.V. would also like to thank the Yukawa Institute for Theoretical Physics at Kyoto University for hospitality during the workshop YITP-T-18-04 "New Frontiers in String Theory 2018" while this work was in progress.

\appendix

\section{Explicit representation matrices} \label{matrea}
For general $N$, we can take the explicit representation of \cite{Castro:2011iw} with all odd spin generators multiplied by $\sqrt{\s}$. Concretely, for
$N=2$:
\be
L_{-1} = \left(
\begin{array}{cc}
 0 & 1 \\
 0 & 0 \\
\end{array}
\right), \qquad L_0= \left(
\begin{array}{cc}
 \frac{1}{2} & 0 \\
 0 & -\frac{1}{2} \\
\end{array}
\right), \qquad L_1 =\left(
\begin{array}{cc}
 0 & 0 \\
 -1 & 0 \\
\end{array}
\right) \, .
\ee
For $N=3$:
\be
L_{-1}=
\left(
\begin{array}{ccc}
 0 & \sqrt{2} & 0 \\
 0 & 0 & \sqrt{2} \\
 0 & 0 & 0 \\
\end{array}
\right),\ L_0 = \left(
\begin{array}{ccc}
 1 & 0 & 0 \\
 0 & 0 & 0 \\
 0 & 0 & -1 \\
\end{array}
\right),\  L_1=\left(
\begin{array}{ccc}
 0 & 0 & 0 \\
 -\sqrt{2} & 0 & 0 \\
 0 & -\sqrt{2} & 0 \\
\end{array}
\right),
\ee
\be 
W_{-2} = \sqrt{\s} \left(
\begin{array}{ccc}
 0 & 0 & 2 \\
 0 & 0 & 0 \\
 0 & 0 & 0 \\
\end{array}
\right),\ W_{-1} = \sqrt{\s}\left(
\begin{array}{ccc}
 0 & \frac{1}{\sqrt{2}} & 0 \\
 0 & 0 & -\frac{1}{\sqrt{2}} \\
 0 & 0 & 0 \\
\end{array}
\right),\ \ W_{0} = \sqrt{\s}\left(
\begin{array}{ccc}
 \frac{1}{3} & 0 & 0 \\
 0 & -\frac{2}{3} & 0 \\
 0 & 0 & \frac{1}{3} \\
\end{array}
\right),
\ee
\be 
W_{1} = \sqrt{\s} \left(
\begin{array}{ccc}
 0 & 0 & 0 \\
 -\frac{1}{\sqrt{2}} & 0 & 0 \\
 0 & \frac{1}{\sqrt{2}} & 0 \\
\end{array}
\right),\ W_{2} = \sqrt{\s}\left(
\begin{array}{ccc}
 0 & 0 & 0 \\
 0 & 0 & 0 \\
 2 & 0 & 0 \\
\end{array}
\right)\, .
\ee 

\section{Chevalley basis for $\mathcal{A}_{N-1}$}\label{connectionbasis}

Here we write the matrix basis we used to write down the form of the Toda connection for general $N$, \eqref{slnconnection2}. The $N\times N$ matrices can be written down in terms of Kronecker's delta
\begin{equation}
    \left(H_i\right)_{jk}=\delta_{ij}\delta_{ik}-\delta_{i+1,j}\delta_{i+1,k}\, , \quad \left(E_i^+\right)=\delta_{ij}\delta_{i+1,k} \, ,\quad \left(E_i^-\right)=\delta_{i+1,j}\delta_{ik} \, ,
\end{equation}
where $i=1,\ldots,N-1$.
Specifically for $N=3$ we have
\begin{equation}\label{Hsl3}
 H_1 =\begin{pmatrix}
1 & 0 & 0 \\
0 & -1 & 0 \\
0 & 0 & 0
\end{pmatrix} \, ,\quad
 H_2 = \begin{pmatrix}
0 & 0 & 0 \\
0 & 1 & 0\\
0 & 0 & -1
\end{pmatrix}\, ,
\end{equation}
\begin{equation}
 E_1^+ =\begin{pmatrix}
0 & 1 & 0 \\
0 & 0 & 0 \\
0 & 0 & 0
\end{pmatrix} \, ,\quad
 E_1^- = \begin{pmatrix}
0 & 0 & 0 \\
1 & 0 & 0\\
0 & 0 & 0
\end{pmatrix}\, ,
\end{equation}
\begin{equation}
 E_2^+ =\begin{pmatrix}
0 & 0 & 0 \\
0 & 0 & 1 \\
0 & 0 & 0
\end{pmatrix} \, ,\quad
 E_2^- = \begin{pmatrix}
0 & 0 & 0 \\
0 & 0 & 0\\
0 & 1 & 0
\end{pmatrix}\, .
\end{equation}
%

\section{Aspects of Toda theory}\label{TodaTheory}
In this appendix we review general aspects of the $\mathcal{A}_{N-1}$ Toda theory that we use in the main text. For the case of real Toda fields, this is material known in the literature and can be found in various sources including \cite{BabylonTalon,Bilal:1988jf,Fateev:2007ab}. Here we will start by assuming the Toda fields to be complex and we will later specify two different reality conditions on them.
The system of Toda equations consists of $N-1$ fields $\phi_i$ satisfying 
\begin{equation}\label{ANTodaEqns}
    \begin{matrix}
    \partial\bar{\partial}\phi_1+ e^{-2\phi_1+\phi_2}=0 \, ,\\
    \partial\bar{\partial}\phi_2+ e^{-2\phi_2+\phi_1+\phi_3}=0 \, ,\\
    \vdots\\
    \partial\bar{\partial}\phi_i+ e^{-2\phi_i+\phi_{i-1}+\phi_{i+1}}=0 \, ,\\
    \vdots\\
    \partial\bar{\partial}\phi_{N-1}+ e^{-2\phi_{N-1}+\phi_{N-2}}=0\, .
    \end{matrix}
\end{equation}
We observe that the above system has a $\mathbb{Z}_2$ symmetry under exchanging $\phi_i\longleftrightarrow\phi_{N-i}$. Also if we know one of the Toda fields, let's say $\ex^{\phi_1}$, we can find the rest by substituting in the Toda equations and solving them iteratively. The field $\ex^{\phi_1}$ can be found by solving an $N$-th order associated ODE and the related monodromy problem.

\subsection{The associated ODE}
Two expressions for the $N$-th order ODE are
\begin{equation}\label{ANODES}
    \prod_{i=1}^N\left(\partial + J_i\right)\xi  = \left(\partial^N + \sum_{i=2}^NU^{(i)}\partial^{N-i}\right)\xi=0\, ,
\end{equation}
where we have assumed $U^{(1)}=\sum_{i=1}^N J_i=0$. The relation between $J_i$ and $U^{(i)}$ can be seen by expanding the above equation. Specifically for the $\mathcal{A}_1$ Toda (Liouville) we have 
\begin{equation}\label{A2UJ}
    U^{(2)}\equiv T = -\partial J -J^2\, ,
\end{equation}
where $J_1=-J_2\equiv J$ and $T$ is the Liouville stress tensor.\\
For the case of the $\mathcal{A}_2$ Toda we have
\begin{gather}\label{A3UJ}
U^{(3)} = J_1  \partial J_3+ J_1  J_2 J_3 +  \partial^2 J_3 + \partial J_2 J_3 +  J_2 \partial J_3 \, ,
\\
U^{(2)}= J_1 J_2 + J_2 J_3 + J_1 J_3 + \partial J_3 - \partial J_1\, .
\end{gather}
Upon identifying 
\begin{equation}\label{Jphi1holo}
J_i \equiv \left(\partial\phi_{N-j}H^j\right)_{ii}\, ,
\end{equation}
and provided the Toda equations are satisfied we get that the functions $U^{(s)}$ correspond to $N-1$ holomorphic currents
\begin{equation}
    \bar{\partial} U^{(s)}_1 = 0 \, , \quad s=2,\ldots,N\, .
\end{equation}
Similarly we define $N-1$ antiholomorphic currents
\begin{equation}
    \partial V^{(s)}_1 = 0 \, , \quad s=2,\ldots,N\, ,
\end{equation}
which are constructed out of $\bar{\partial}^k\phi_i$. Specifically for the case of $sl(3)$ we have
\begin{equation}\label{1Ucurrents}
    \begin{split}
        & U^{(2)}_1 = -\partial^2\phi_1 - \partial^2\phi_2 - (\partial\phi_1)^2 - (\partial\phi_2)^2 + \partial\phi_1\partial\phi_2 \, ,\\
        & U^{(3)}_1 = \partial^3\phi_1 + \partial\phi_1\left(-2\partial^2\phi_1 + \partial^2\phi_2 - \partial\phi_1\partial\phi_2 + (\partial\phi_2)^2  \right)\, ,
    \end{split}
\end{equation}
and
\begin{equation}\label{1Vcurrents}
    \begin{split}
        & V^{(2)}_1 = -\bar{\partial}^2\phi_1 - \bar{\partial}^2\phi_2 - (\bar{\partial}\phi_1)^2 - (\bar{\partial}\phi_2)^2 + \bar{\partial}\phi_1\bar{\partial}\phi_2 \, ,\\
        & V^{(3)}_1 = \bar{\partial}^3\phi_1 + \bar{\partial}\phi_1\left(-2\bar{\partial}^2\phi_1 + \bar{\partial}^2\phi_2 - \bar{\partial}\phi_1\bar{\partial}\phi_2 + (\bar{\partial}\phi_2)^2  \right)\, .
    \end{split}
\end{equation}
The currents $V^{(s)}$ can be taken from the currents $U^{(s)}$ simply by replacing $\partial$ with $\bar{\partial}$.
Then we have two ODEs, one for the holomorphic and one for antiholomorphic currents, which are both satisfied by $\ex^{\phi_1}$
\begin{equation}\label{TodaANU1}
\begin{split}
& \left(\partial^N + {U}^{(2)}_1\partial^{N-2}+\ldots +U^{(N)}_1\right)e^{\phi_{1}}=0 \, , \\
& \left(\bar{\partial}^N + V^{(2)}_1\bar{\partial}^{N-2}+\ldots +V^{(N)}_1\right)e^{\phi_1}=0\, .
 \end{split}
\end{equation}
Because of the $\mathbb{Z}_2$ symmetry of the Toda system we can also make the identification
\begin{equation}\label{Jphi2holo}
J_i \equiv  \left(\partial\phi_jH^j\right)_{ii}\, .
\end{equation}
Then we get a second set of holomorphic and anti-holomorphic currents 
\begin{equation}\label{U2holoantiholo}
    \bar{\partial} U^{(s)}_{2} = 0 \, , \quad \partial V^{(s)}_{2} = 0 \, , \quad s=2,\ldots,N\, ,
\end{equation}
and a second set of ODEs that are being satisfied by $\ex^{\phi_{N-1}}$
\begin{equation}\label{TodaANU2}
\begin{split}
 & \left(\partial^N + {U}^{(2)}_2\partial^{N-2}+\ldots +U^{(N)}_2\right)e^{\phi_{N-1}}=0 \, , \\
&  \left(\bar{\partial}^N + V^{(2)}_2\bar{\partial}^{N-2}+\ldots +V^{(N)}_2\right)e^{\phi_{N-1}}=0\, .
\end{split}
\end{equation}
In other words we can get \eqref{TodaANU2} from \eqref{TodaANU1} be exchanging $\phi_i \leftrightarrow \phi_{N-i}$.

\subsection{Primary currents}\label{PrimaryCurrentsGeneral}

One can check that under a conformal mapping $z\rightarrow f(z)$ the Toda system \eqref{ANTodaEqns} is invariant provided the fields $\phi_i$ transform as
\begin{equation}\label{phiconftrans}
    \phi_i \rightarrow \phi_i - \frac{q_i}{2}\log\left(\partial f(z)\right) - \frac{q_i}{2}\log\left(\bar{\partial} \bar{f}(\bar{z})\right)\, .
\end{equation}
The constants $q_i$ are the proportionality constants $\phi_i=q_i\phi_{grav}$ that appear  after restricting to the gravity subsector of the theory as 
\begin{equation}\label{phigravsubsector}
    \phi_jH^j = 2\phi_{grav}L_0\, .
\end{equation}
From this we immediately deduce $q_1=q_{N-1}=N-1$. For example for $N=2$ we have $q=1$ and for $N=3$ we have $q_1=q_2=2$.

Out of the currents $U^{(s)}_{1,2}$ and their derivatives one can construct currents $\mathcal{W}^{(s)}_{1,2}$ that transform as primaries under the above transformation
\begin{equation}\label{primarytrans}
    \mathcal{W}^{(s)}_{1,2} \rightarrow \left(\partial f(z)\right)^s \mathcal{W}^{(s)}_{1,2}\, , \quad s=3,\ldots N \, .
\end{equation}
For $s=2$ the current transforms as a quasi-primary
\begin{equation}\label{quasiprimarytrans}
    U^{(2)}_{1,2}\equiv \mathcal{W}^{(2)}_{1,2}\rightarrow \left(\partial f(z)\right)^2 \mathcal{W}^{(2)}_{1,2} + \beta_N S\left(f(z),z\right)\, ,
\end{equation}
where $\beta_N$ is a constant and 
\begin{equation}\label{Schwarzian}
    S\left(f(z),z\right) \equiv \frac{\partial^3f(z)}{\partial f(z)} - \frac{3}{2}\left( \frac{\partial^2f(z)}{\partial f(z)} \right)^2\, ,
\end{equation}
is the Schwarzian derivative.\\
Specifically for the $\textrm{sl}(2)$ case we have $\beta_2=\frac{1}{2}$ while for $\textrm{sl}(3)$ we have $\beta_3=2$ and 
\begin{equation}\label{N3W3}
    \mathcal{W}^{(3)}_{1,2} = U^{(3)}_{1,2} - \frac{1}{2}\partial\mathcal{W}^{(2)}_{1,2} \, .
\end{equation}
From \eqref{N3W3} and \eqref{1Ucurrents} we get
\bea \label{currentsTWapp}
\calw^{(2)}_1 (z) &=& -(\pa \f_1 )^2 -(\pa \f_2 )^2 + \pa \f_1 \pa \f_2 - \pa^2 \f_1 - \pa^2 \f_2 \\ 
\calw_1^{(3)} (z) &=& - (\pa \f_1 )^2 \pa \f_2 +  (\pa \f_2 )^2 \pa \f_1  -  \pa^2 \f_1  \pa \f_1 +  \pa^2 \f_2 \pa \f_2\nonu 
&&+ \half \left( - \pa^2 \f_1  \pa \f_2 +  \pa^2 \f_2  \pa \f_1 - \pa^3 \f_1 + \pa^3 \f_2 \right) \, .
\eea
Thus for the case of $\textrm{sl}(3)$ we observe that
\begin{equation}
    \calw^{(2)}_{1}=\calw^{(2)}_{2}\, , \quad \calw^{(3)}_{1}= - \calw^{(3)}_{2} \, .
\end{equation}
In general for the primary currents we have the relation
\begin{equation}\label{currents12relation}
    \mathcal{W}^{(s)}_1 = \left(-1\right)^s \mathcal{W}^{(s)}_2\, ,\quad s=2,\ldots,N\, .
\end{equation}
Thus unless otherwise specified for the primary currents we will drop the bottom index and write
\begin{equation}\label{currents1relation}
    \mathcal{W}^{(s)} \equiv \mathcal{W}^{(s)}_1\, ,\quad s=2,\ldots,N\, .
\end{equation}
Similarly out of the currents $V^{(s)}_{1,2}$ and their derivatives one can construct currents $\mathcal{W}^{(s)}_{b}$ that transform as primaries under transformation \eqref{phiconftrans}. Now we have already assumed \eqref{currents1relation} and the sub-index $b$ simply denotes the primary currents constructed out of $V^{(s)}$ instead of $U{(s)}$.

\subsection{Properties of the solutions}
As we alluded in the beginning of this section to fully solve the Toda system we only need to know one of the Toda fields which can be found by satisfying an associated ODE. For the field $\phi_1$ we showed that the holomorphic and antiholomorphic ODEs are given by \eqref{TodaANU1}.
Thus we can write
\begin{equation}\label{phi1decomposition}
\ex^{\phi_1} = \sum_{i=1}^N\psi_i\left(z\right)\tilde{\psi}^i\left(\bar{z}\right)\, ,
\end{equation}
where $\psi_i$ and $\tilde{\psi}^i$ are independent holomorphic and anti-holomorphic solutions of \eqref{TodaANU1}
\begin{equation}\label{ODEpsi}
\left(\partial^N + U^{(2)}_1\bar{\partial}^{N-2}+\ldots +U^{(N)}_1\right)\psi_i(z)=0 \, , 
\end{equation}
\begin{equation}\label{ODEpsitilde}
\left(\bar{\partial}^N + V^{(2)}_1\bar{\partial}^{N-2}+\ldots +V^{(N)}_1\right)\tilde{\psi}^i(\bar{z})=0\, .
\end{equation}
Similarly for the field $\phi_{N-1}$ we have
\begin{equation}
    \ex^{\phi_{N-1}} = \sum_{i=1}^N\chi^i\left(z\right)\tilde{\chi}_i\left(\bar{z}\right)\, ,
\end{equation}
where $\chi^i$ and $\tilde{\chi}_i$ are independent holomorphic and anti-holomorphic solutions of \eqref{TodaANU2}
\begin{equation}\label{ODEchi}
\left(\partial^N + U^{(2)}_2\bar{\partial}^{N-2}+\ldots +U^{(N)}_2\right)\chi^i(z)=0 \, , 
\end{equation}
\begin{equation}\label{ODEchitilde}
\left(\bar{\partial}^N + V^{(2)}_2\bar{\partial}^{N-2}+\ldots +V^{(N)}_2\right)\tilde{\chi}_i(\bar{z})=0\, .
\end{equation}
To further analyze the solutions we will need to impose a reality condition on the Toda fields. Here we will consider two different reality conditions. In the first case we will take the Toda fields to be real. In the second case we will take the Toda fields to be complex conjugate of each other such that $\phi_i = \bar{\phi}_{N-i}$.

\subsubsection{Real Toda fields}
For convenience we arrange the solutions into a column vector $\Psi=\left(\psi_1, \ldots\psi_N\right)^T$. When the Toda fields are real we observe that
\begin{equation}
    V_i^{(s)} = \bar{U}_i^{(s)}\, .
\end{equation}
Thus $\bar{\Psi}(\bar{z})$ and $\tilde{\Psi}$ solve the same ODE. Then in general, $\tilde{\psi}^i$ will be linear combinations of $\bar{\psi}_i$
\begin{equation}\label{tildebarpsi}
\widetilde{\Psi} = \Lambda^T \bar{\Psi}\, ,
\end{equation}
where $\Lambda \in \textrm{GL}\left(N,\mathbb{C}\right)$ is a constant matrix such that
\begin{equation} \label{psiLpsi}
\ex^{\phi_1} = \Psi^{\dagger}\left(\bar{z}\right)\Lambda\Psi\left(z\right)\, .
\end{equation}
Taking the conjugate transpose of the above and demanding $\ex^{\phi_1}$ to be real we get the condition
\begin{equation}\label{unitarycondition}
\Lambda^{\dagger} = \Lambda\, .
\end{equation}
As we mentioned before by knowing the solution for $\ex^{\phi_1}$ and by substituting in the Toda equations we can iteratively obtain all the $\ex^{\phi_i}$ with $i=1, \ldots, N-1$. To perform this procedure we actually need only $N-2$ from the total of $N-1$ Toda equations. Substituting all the $\ex^{\phi_i}$ in the last Toda equation we obtain the condition
\begin{equation}\label{detproduct}
W_{\psi}\,\textrm{det}\Lambda\,\overline{W}_{\psi} = \left(-1\right)^{\lfloor \frac{N}{2} \rfloor}\, ,
\end{equation}
where $W_{\psi}$ is the Wronskian. Also because $U^{(1)} = \bar{U}^{(1)} = 0$, since $\sum_i J_i =0$, we have that the Wronskian is constant and we can set it equal to one. Therefore \eqref{detproduct} becomes
\begin{equation}\label{scondition}
\textrm{det}\Lambda = \left(-1\right)^{\lfloor \frac{N}{2} \rfloor}\, .
\end{equation}
To fully specify $\Lambda$ we will consider a specific solution of \eqref{ODEpsi} which we will describe in the next subsection.\\
Similarly from the ODEs \eqref{TodaANU2} we have
\begin{equation}\label{XiLambdaXi}
\ex^{\phi_{N-1}} = X^{\dagger}\Lambda_2X\, .
\end{equation}
Starting from $\ex^{\phi_1}$ and by iteratively solving the Toda equations we deduce
\begin{equation}\label{chiMinorpsi}
\chi_i = M^{\{N,i\}}_{W_{\Psi}}\, , \quad i=1,\ldots, N\, ,
\end{equation}
and $\Lambda_2=\Lambda$, where $M^{\{N,i\}}_{W_{\Psi}}$ is the $\{N,i\}$ minor of the Wronskian with respect to $\Psi$.

\subsubsection{Complex conjugate Toda fields $\phi_i = \bar{\phi}_{N-i}$}
In this case the number of independent Toda equations reduces by half as the second half of Toda equations are the complex conjugate of the first half. This can be seen as a consequence of the $\mathbb{Z}_2$ symmetry of the system of Toda equations under exchanging $\phi_i \longleftrightarrow \phi_{N-i}$. In the case of an odd number of Toda fields the middle field is real. When we use this reality condition we observe that
\begin{equation}
    \bar{U}_1^{(s)} = V_2^{(s)}\, .
\end{equation}
Thus $\tilde{\Psi}(\bar{z})$ and $\bar{X}(\bar{z})$ solve the same ODE. The same holds for the pair $\tilde{X}(\bar{z})$ and $\bar{\Psi}(\bar{z})$. Consequently we have
\begin{equation}
    \tilde{\Psi} = N_1^T\bar{X} \, , \quad \tilde{X} = N_2^T\bar{\Psi}\, ,
\end{equation}
where $N_{1,2} \in \textrm{GL}\left(N,\mathbb{C}\right)$ are constant matrices such that
\begin{equation}
    \ex^{\phi_1} = X^{\dagger}N_1\Psi \, , \quad \ex^{\phi_{N-1}} = \Psi^{\dagger}N_2X\, .
\end{equation}
Since we have $\phi_1 = \bar{\phi}_{N-1}$ we deduce that $N_1^{\dagger} = N_2$. By substituting in the Toda equations we find that
\begin{equation}
    N_1=N_2 = \mathbb{I}\, ,
\end{equation}
\begin{equation}
    \psi_a = (-1)^{\lfloor \frac{N-1}{2}\rfloor}\im^{N-1}\epsilon_{ab_1\ldots b_{N-1}}\chi^{b_1}\partial\chi^{b_2}\ldots \partial^{N-2}\chi^{b_{N-1}}\, ,
\end{equation}
\begin{equation}
    \chi^a = \left( (-1)^{\lceil\frac{N}{2}\rceil} \im \right)^{N-1}\epsilon^{ab_1\ldots b_{N-1}}\psi_{b_1}\partial\psi_{b_2}\ldots\partial^{N-2}\psi_{b_{N-1}}\, .
\end{equation}
Thus we can write
\begin{equation}\label{phi1slnR}
    \ex^{\phi_1} = \left( (-1)^{\lceil\frac{N}{2}\rceil} (-\im) \right)^{N-1}\epsilon^{ab_1\ldots b_{N-1}}\psi_a\bar{\psi}_{b_1}\bar{\partial}\bar{\psi}_{b_2}\ldots\bar{\partial}^{N-2}\bar{\psi}_{b_{N-1}}\, .
\end{equation}
%

\subsection{A simple solution}\label{simplesolution}
A simple solution to the associated ODE can be found if we set all the currents to zero, $U^{(s)}_i=0$ and $V^{(s)}_i=0$. Then the $\psi_i$ should be given by linear combinations of $z^k$ with $k=0,1,\ldots ,N-1$. Here we present these solutions for the two reality conditions on the Toda fields

\subsubsection{For real Toda fields}
We have
\begin{equation}\label{SUzsolution}
    \Psi = \frac{1}{A^{\frac{1}{N}}}\left(1, \sqrt{(N-1)}z,\ldots, \sqrt{\binom{N-1}{k}}z^k, \ldots, z^{N-1}\right)^T  \, ,  
\end{equation}
where 
\begin{equation}
    A=\prod_{k=0}^{N-1}k!\sqrt{\binom{N-1}{k}}\, ,
\end{equation}
is a normalization constant chosen such that the Wronskian is equal to one.
To have proper boundary conditions on the disk we want as $|z|\rightarrow 1$
\begin{equation}
    \ex^{\phi_1} \sim \left(1-|z|^2\right)^{q_1} = \left(1-|z|^2\right)^{N-1}\, .
\end{equation}
Then \eqref{psiLpsi} and \eqref{scondition}, together with \eqref{SUzsolution} imply
\begin{equation}
    \Lambda = \textrm{diag}\left(1,-1,1,-1,\ldots\right)\, .
\end{equation}
Then we have
\begin{equation}
    \ex^{\phi_1} =  \frac{1}{A^{\frac{2}{N}}}\left(1-|z|^2\right)^{N-1}=\frac{1}{\left(N-1\right)!}\left(1-|z|^2\right)^{N-1}\, ,
\end{equation}
from which we observe that the solution \eqref{SUzsolution} also specifies the zeroth order term in the expansion of $\phi_1$ 
as
\begin{equation}
    f_0=-\log\left(N-1\right)!\, ,
\end{equation}
For the reflection property of $\psi_i$
\begin{equation}
    \Psi = Sz^{N-1}\bar{X}\left(\frac{1}{z}\right)\, ,
\end{equation}
we find the proportionality matrix $S$ to be
\begin{equation} \label{SNmatrix}
    S=(-1)^{\lfloor\frac{N}{2}\rfloor}\frac{B}{A^{\frac{N-2}{N}}}\mathbb{I} = (-1)^{\lfloor\frac{N}{2}\rfloor}\mathbb{I}\, ,
\end{equation}
where
\begin{equation}
    B=\prod_{k=0}^{N-2}k!\sqrt{\binom{N-1}{k}}\, .
\end{equation}
%

\subsubsection{For complex conjugate Toda fields}
We have
\begin{equation}\label{SLNRzsolution}
    \Psi = \frac{1}{A^{\frac{1}{N}}}\left(1+\im z^{N-1},z+\im z^{N-2},\ldots,z^k+\im z^{N-k-1}\,\ldots , z^{N-1} + \im      \right)^T\, ,
\end{equation}
where
\begin{equation}
    A = \left(1+\im\right)^{\lceil\frac{N}{2}\rceil}\left(1-\im\right)^{\lfloor\frac{N}{2}\rfloor}\prod_{k=1}^{N-1}k!\, ,
\end{equation}
is chosen again such that the Wronskian is equal to one. By using \eqref{SLNRzsolution} together with \eqref{phi1slnR} we find for $N$ being odd
\begin{equation}
    \ex^{\phi_1} = \frac{(-1)^{\frac{2}{N}}}{(N-1)!}\left(1-|z|^2\right)^{N-1}\, ,
\end{equation}
while for $N$ being even
\begin{equation}
    \ex^{\phi_1} = \frac{1}{(N-1)!}\left(1-|z|^2\right)^{N-1}\, .
\end{equation}
The solution specifies again the zeroth order term in the expansion of $\phi_1$ which is now complex for $N$ being odd
\begin{equation}
    f_0 = -\log(N-1)! - n\frac{2\pi\im}{N}\, ,
\end{equation}
where $n$ is an integer.
For the reflection property of $\psi_i$
\begin{equation}
    \Psi = \tilde{S}z^{N-1}\bar{\Psi}\left(\frac{1}{z}\right)\, ,
\end{equation}
we find the proportionality matrix to be
\begin{equation} \label{tildeSNmatrix}
    \tilde{S} = \im \left(\frac{\bar{A}}{A}\right)^{\frac{1}{N}}\mathbb{I}\, .
\end{equation}
%

\section{Coupling higher spin particles via Wilson lines}\label{WilsonParticles}
The dynamics of massive point particles coupled to gravity in three dimensions is captured, in the Chern-Simons formulation, by  Wilson lines. This connection was initially established for the case of asymptotically flat gravity \cite{Witten:1989sx,Carlip:1989nz}. In the context of entanglement entropy it has also been extended to AdS$_3$ both for  higher spin theories \cite{Ammon:2013hba, Castro:2014mza} and for spinning particles \cite{Castro:2014tta}. Here we review this construction in for the situation relevant to us, namely that we consider the Lorenzian higher spin theories with independent gauge fields $A, \bar A$ taking values in $\textrm{sl}(N, \RR)$ or $\textrm{su}(\lfloor \frac{N}{2} \rfloor , \lceil\frac{N}{2}\rceil)$, and that we restrict attention to `chiral' point-particle sources which couple only to $A$.

\subsection{Wilson Lines}\label{WilsonLines}
We start from the higher spin theory described by two Chern-Simons fields  $A, \bar A$  taking values  in $\textrm{sl}(N, \RR)$ or $\textrm{su}(\lfloor \frac{N}{2} \rfloor , \lceil\frac{N}{2}\rceil)$.
The most general Wilson line we can add to this theory is of the form  $W_R (C) \bar W_{\bar R} (C)$, where
 \be 
W_R (C)  = \tr_R \calp \exp \int_C A , \qquad \bar W_{\bar R} (C)  = \tr_{\bar R} \calp \exp \int_C \bar A  \, .
\label{WLapp} \ee
It depends both on the choice of the curve $C$ and on two representations $R, \bar R $ of the gauge group. 
It is natural to expect that, if we take $R$ and $\bar R$ to be unitary irreducible representations, the Wilson line describes a coupling of a point particle to the higher spin theory, with $R$ and $\bar R$ carrying  the information about the physical properties point particle such as mass, spin and higher spin charges. This interpretation was made precise in
\cite{Ammon:2013hba}, see also \cite{Castro:2014tta}, \cite{Castro:2018srf}.
Restricting attention to representations where the energy $P_0 = L_0 - \bar L_0$ is bounded below, we are led to the infinite-dimensional, lowest-weight representations. These are built on a lowest weight or primary
state $|h, \vec w\rangle=|h,w_3,...,w_N\rangle$ satisfying
\begin{equation}\label{hwstate}
\begin{split}
&  L_0|h,\vec w\rangle = h|h, \vec w\rangle \, , \qquad  L_1|h, \vec w\rangle = 0 \, , \\
&  W^{(s)}_0|h, \vec w\rangle = w_s|h, \vec w\rangle \, , \qquad W^{(s)}_j|h, \vec w\rangle = 0 \, , \quad j=1,\ldots, s-1\, ,
\end{split}
\end{equation}
where $s=3,\ldots ,N$. The primary state is annihilated by the lowering operators and descendant states are created by acting with the raising operators $L_{-1}$ and $W^{(s)}_{-j}$. The physical properties of the particle are encoded in the $2 \times (N-1)$ independent Casimir invariants of the representation $R$ and $\bar R$. These can be expressed in terms of the primary weights $(h, \vec w)$ (and their right-moving cousins) as we will work out below for $N=2,3$. In this work we are be interested in the subclass of particles where 
\be \bar R =1\, , \ee 
is the singlet representation; so from now on we will deal only with the left-moving Wilson line
$W_R (C)$.

A useful representation of the  Wilson line  is obtained by interpreting $R$ as the Hilbert space of an auxiliary quantum mechanical system that lives on the Wilson line. The auxiliary quantum system is described by a field $U$ taking values in the  gauge group ($\textrm{SL}(N,\RR) $ or  $\textrm{SU}(\lfloor \frac{N}{2} \rfloor , \lceil\frac{N}{2}\rceil)$) and it's conjugate momentum $P$ taking values in the Lie algebra. The dynamics of $U, P$ is picked so that upon quantisation the Hilbert space of the system will be the representation $\mathcal{R}$. Then the trace over $R$ is replaced by a path integral
\begin{equation}
W_{R}(C) = \int\mathcal{D}U e^{ i S(U;A)_{R,C}}\label{PI}\, ,
\end{equation}
where $ S(U;A)_{R,C}$ is a first-order action of the form \cite{Castro:2014tta}
\begin{equation}\label{leftaction}
  S(U;A)_{R,C} = \int_C ds\left[ \tr\left(P D_{s}UU^{-1}\right)  + \lambda^{(2)}\left(\tr\left(P^2\right)+c_R^{(2)}\right) + \ldots + \lambda^{(N)}\left(\tr\left(P^N\right)+c_R^{(N)}\right)\right]\, ,
\end{equation}
and
\begin{equation}
D_{s}U=\partial_s U +A_s U\, ,\quad A_s=A_{\mu}\frac{dx^{\mu}}{ds}\, .
\end{equation}
Here $A_s$ denotes the pullback of $A$ to the world-line $C$ and $P$ is a canonically conjugate momentum to $U$ and takes values in the Lie algebra $\textrm{sl}(N,\RR) $ resp.  $\textrm{su}(\lfloor \frac{N}{2} \rfloor , \lceil\frac{N}{2}\rceil)$.  We note that the $\l^{(i)}$ are Lagrange multipliers which fix the trace invariants of $P$ in terms of the Casimir invariants $c_R^{(i)}$ of the representation $R$.
We refer to \cite{Castro:2018srf} for more details on the equivalence between (\ref{WLapp}) and (\ref{leftaction}). 

The action (\ref{leftaction}) is invariant under the gauge symmetry $A \to \L ( A + d ) \L^{-1} $ under which the worldline fields  transform as
\be
U \to \L U, \qquad P \to \L P \L^{-1} \, ,
\ee
where in this formula $\L =\L (x^\m(s))$ is  pulled back to the worldline.
In the above action we take the  trace ``$\tr$"  to be normalized as in the $N$-dimensional representation.
This defines the Killing forms
\begin{equation}
h_{a_1...a_m} = \tr\left(T_{\small(a_1}...T_{a_m\small)}\right)\, , \quad m = 2, \ldots, N\, ,
\end{equation}
where $T_a$ are the generators of $\textrm{sl}(N,\RR)$ resp.   $\textrm{su}(\lfloor \frac{N}{2} \rfloor , \lceil\frac{N}{2}\rceil)$.
The Casimir operators are given by
\begin{equation}
C^{(m)} = h^{a_1...a_m}T_{a_1}...T_{a_m}\, ,\label{Casdef}
\end{equation}
and the $c^{(m)}_{R}$ are their values in the representation $R$. Specifically for the momentum we have
\begin{equation}
    \tr\left(P^m\right) = h_{a_1\ldots a_m}P^{a_1}\ldots P^{a_m}\, ,
\end{equation}
where $P=P^aT_a$. We will be interested in the  regime of the parameters $(h, \vec w)$ where the path integral (\ref{PI}) is well approximated   by its saddle point value; in this regime we have to find solutions the equations following from the total action
\begin{equation}
S= S_{CS}(A) - S_{CS}(\bar{A}) +S(U;A)_{R,C}\, .\label{totaction}
\end{equation}
%

\subsection{Equations of motion}\label{WilsonEoms}
From the above action we derive  following equations of motion for the connections
\be 
 F_{\mu\nu}= - \frac{2\pi}{k}\epsilon_{\mu\nu\rho}\int_C ds \frac{dx^{\rho}}{ds} P\delta^{(3)}\left(x - x(s)\right), \qquad 
\bar{F}_{\mu\nu}=0\, ,\label{Feq}
\ee
where $k$ is related to the central charge and Newton's constant through
\begin{equation}\label{kgeneral}
c=12k\epsilon_N=\frac{3}{2G}\, ,
\end{equation}
where
\begin{equation}
\epsilon_N =\textrm{Tr}[L_0 L_0]=\frac{N\left(N^2-1\right)}{12}\, .
\end{equation}
The equation from varying $U$ is
\begin{equation} \label{Ueom}
 \partial_ sP + \left[P,\partial_sU U^{-1}\right] = 0 \, ,
\end{equation}
and from varying the momentum we obtain
\begin{equation} \label{Peom}
\frac{1}{2}D_{s}UU^{-1} + 2\lambda^{(2)}P + 3\lambda^{(3)}P\times P+...+N\lambda^{(N)}\underbrace{P\times ...\times P}_{N-1}= 0\, ,
\end{equation}
where
\begin{equation}
\underbrace{P\times ... \times P}_{m}=h_{a_1...a_{m+1}}P^{a_1}...P^{a_m}T^{a_{m+1}}\, .
\end{equation}
We also have the constraints coming from the Lagrange multipliers
\begin{equation} \label{slnconstraints}
\tr\left(P^m\right) = - c_R^{(m)},  \quad m=2,...,N\, .
\end{equation}

Let us now simplify the system equations (\ref{Feq},\ref{Ueom},\ref{Peom},\ref{slnconstraints})  for the situation at hand. As we argued and showed explicitly for $N=2$ and $N=3$, we can choose a 
coordinate system $(t,z ,\bar z)$ and a suitable gauge such that $A$ is a Lax connection  for the $\cala_{N-1}$ Toda system. In particular, $A$ is of the form $A_z dz + A_{\bar z} d \bar z$ and  the worldline $C$ the particle moves on has constant $z=z_0$. In this case we can choose the worldline coordinate $s$ such that $t (s)= s$ and we observe that $A_s =0$.
The equations (\ref{Ueom},\ref{Peom}) are then solved by
\be 
\pa_s P=0\, , \qquad U=1\, , \qquad \l^{(m)}=0\, .
\ee
The remaining equation (\ref{Feq}) is then\footnote{Our complex delta-function is normalized as $\frac{\im}{2} \int dz d\bar z  \d^{(2)} (z)=1$ and we have e.g. $\bar \pa \left({1 \over z}\right) = \p \d^{(2)} (z)  $.}
\be
F_{z \bar z} = - {\p \im \over k} P \d^{(2)} (z-z_0)\, ,\label{Feqcompl}
\ee 
where the constant Lie algebra element $P$ is constrained to satisfy  (\ref{slnconstraints}).  Since $A$ is a Lax connection for the $\cala_{N-1}$ Toda system, the above equations reduce to the Toda equations with delta-function sources. In the following two paragraphs we will work the precise coefficients in front of the delta-functions for  $N=2$ and $N=3$.

\subsection{Spin 2 case}\label{spin2WL}
In this case we have (see (\ref{Fgrav}))
\be 
V^{-1} \tilde{F}_{z \bar z} V = - 2 \left( \pa \bar \pa \f + e^{-2  \f}\right) L_0\, ,
\ee
where $V$ is constant element  defined in (\ref{Vdef}). Plugging the latter into  (\ref{Feqcompl}) shows that $V^{-1}\tilde{P} V$ should be proportional to $L_0$:
\be 
V^{-1}\tilde{P} V = - \im 2 \a L_0\, .
\ee
The constant $\a$ is determined by the Casimir constraint (\ref{slnconstraints}). From (\ref{Casdef}) the quadratic Casimir takes the value
\be 
c^{(2)}_R = 2 h(h-1) \approx 2 h^2\, ,
\ee
where in the last approximation we used the fact that the saddle point approximation to (\ref{PI}) is valid for $h\gg 1$. Equation  \eqref{slnconstraints} then becomes
\be
\tr \tilde{P}^2 = - 2\a^2 =  -  2 h^2 \, ,
\ee
so that  (\ref{Feqcompl}) reduces to
\be \pa \bar \pa + e^{-2  \f} ={ \p h\over k} \d^{(2)} (z-z_0) \, .
\ee
Using (\ref{kgeneral}) and generalizing to several point-particle sources leads to
\begin{equation}
    \partial\bar{\partial}\phi + e^{-2\phi} = 4\pi G \sum_i h_i \delta^{(2)}\left(z-z_i,\bar{z}-\bar{z}_i\right)\, ,
\end{equation}
which is the equation  found in \cite{Hulik:2016ifr} using the metric formulation upon identifying $m_i =  h_i$.

\subsection{Spin 3 case}\label{spin3WL}
In this case we have from \eqref{ATsl3}
\be 
V^{-1} \tilde{F}_{z \bar z} V = - \left(\partial\bar{\partial}\phi_1 + \ex^{-2\phi_1 + \phi_2}\right) H_1 - \left(\partial\bar{\partial}\phi_2 + \ex^{-2\phi_2 + \phi_1}\right) H_2 \, ,
\ee
where $H_1, H_2$ are the diagonal matrices \eqref{Hsl3}\footnote{ These are related to $L_0$ and $W_0$ as 
$H_1 = \half L_0 + {3 \over 2 \sqrt{\s}} W_0$, $H_2 = \half L_0 - {3 \over 2 \sqrt{\s}} W_0$.}.
Setting
\be 
V^{-1}\tilde{P} V = -\im  (\a_1 H_1 + \a_2 H_2),
\ee
 (\ref{Feqcompl}) reduces to
 \begin{equation}\label{spin3EomsWL}
\begin{split}
& \partial\bar{\partial}\phi_1 + \ex^{-2\phi_1 + \phi_2} = 16 \p G \alpha_1\delta^{(2)}\left(z-z_0,\bar{z}-\bar{z}_0\right) \, , \\
& \partial\bar{\partial}\phi_2 + \ex^{-2\phi_2 + \phi_1} = 16 \p G \alpha_2\delta^{(2)}\left(z-z_0,\bar{z}-\bar{z}_0\right)\, ,
\end{split}
\end{equation}
where we have used (\ref{kgeneral}). The constants $\a_{1,2}$ are determined by the Casimir constraints (\ref{slnconstraints}). For the values of the Casimirs we find from (\ref{Casdef})
\be 
c^{(2)}_R = \frac{1}{2}h^2 + \frac{3 \s}{2}w^2 + \ldots, \qquad 
c^{(3)}_R = {3 \over 4 \sqrt{\s}} w ( h^2 - \s w^2) + \ldots\, ,
\ee 
where the omitted terms are subleading in the regime where the saddle point approximation to  (\ref{PI})
is valid. The   Casimir constraints (\ref{slnconstraints}) then reduce to 
\begin{equation}\label{spin3constraintsWL}
\begin{split}
& \a_1^2 - \a_1 \a_2 + \a_2^2 =\frac{1}{4}\left( h^2 + 3 \s w^2\right) \, , \\
& \left(\alpha_2-\alpha_1\right)\alpha_1\alpha_2 = { \im \over \sqrt{4\s}}  w\left(h^2 -\s  w^2\right)\, .
\end{split}
\end{equation}
Equations \eqref{spin3EomsWL} easily generalize to many particles by replacing $\alpha_j\rightarrow \sum_i \alpha^{(i)}_j$. Then the resulting equations are equations \eqref{spin3eoms} appearing in the main text.

\slin


\bibliographystyle{ytphys}
\bibliography{ref}
\end{document}